\documentclass[preprint,12pt,authoryear]{elsarticle}

\usepackage{amsmath,amssymb,graphicx,booktabs,subfigure,caption,url,multirow,xcolor,enumitem}

\usepackage{tikz}
\usetikzlibrary{bayesnet}

\journal{Computer Science and Language}

\def\x{{\mathbf x}}

\newcommand{\T}{{\top}}
\DeclareMathOperator{\Norm}{Norm}

\DeclareMathOperator{\onehot}{onehot}

\DeclareMathOperator{\bin}{bin}

\DeclareMathOperator*{\argmin}{argmin}
\newcommand{\red}{\textcolor{red}}
\begin{document}

\begin{frontmatter}



\title{A Speaker Verification Backend with Robust Performance across Conditions}
 \author[icc]{Luciana Ferrer}
 \author[star]{Mitchell McLaren}
 \author[oml]{Niko Br\"ummer}

\address[icc]{Instituto de Investigaci\'on en Ciencias de la Computaci\'on (ICC), \\CONICET-UBA, Argentina}
\address[star]{Speech Technology and Research Lab (StarLab), SRI International, USA}
\address[oml]{Phonexia, South Africa\footnote{During the first part of this author's contribution, he was with \emph{Omilia Conversational Intelligence}.}}

\begin{abstract}
In this paper, we address the problem of speaker verification in conditions unseen or unknown during development. A standard method for speaker verification consists of extracting speaker embeddings with a deep neural network and processing them through a backend composed of probabilistic linear discriminant analysis (PLDA) and global logistic regression score calibration. This method is known to result in systems that work poorly on conditions different from those used to train the calibration model. We propose to modify the standard backend, introducing an adaptive calibrator that uses duration and other automatically extracted side-information to adapt to the conditions of the inputs. The backend is trained discriminatively to optimize binary cross-entropy. When trained on a number of diverse datasets that are labeled only with respect to speaker, the proposed backend consistently and, in some cases, dramatically improves calibration, compared to the standard PLDA approach, on a number of held-out datasets, some of which are markedly different from the training data. Discrimination performance is also consistently improved. We show that joint training of the PLDA and the adaptive calibrator is essential --- the same benefits cannot be achieved when freezing PLDA and fine-tuning the calibrator. To our knowledge, the results in this paper are the first evidence in the literature that it is possible to develop a speaker verification system with robust out-of-the-box performance on a large variety of conditions.
\end{abstract}



\begin{keyword}
Speaker Verification \sep 
Probabilistic Linear Discriminant Analysis \sep
Robust Calibration 
\end{keyword}

\end{frontmatter}


\section{Introduction}
\def\Dset{\mathcal{D}}
\label{sec:intro}

The task of speaker verification (SV) is to determine whether two sets of speech samples originate from a common speaker, or from two different speakers. SV systems output a score for each comparison --- usually called a \emph{trial} --- which can be used in different ways depending on the application. 
In some cases, like in surveillance applications where the goal is to search for certain target speakers within a large database of speech samples, the scores can be used to obtain a sorted list of potential matches, from most to least likely. This list can then be reviewed by a person to make the final decisions. In other cases, like in authentication scenarios where the system is used to automatically verify whether a person is who they claim to be, a hard decision has to be made for each comparison. This means that the system has to apply a threshold to the scores so that only trials with a score above that threshold are declared same-speaker trials. 

The selection of such a threshold is not a trivial matter. A system with excellent discrimination between same- and different-speaker trials could nevertheless produce extremely poor decisions, on average, if the wrong threshold is used. If a data set that matches the conditions\footnote{What do we mean by \emph{conditions}? They are a collection of (often hidden) factors that differ between different speech samples and that can have the unfortunate effect of also shifting the distributions of speaker verification scores. See section~\ref{sec:condition} below for a more detailed discussion. } of the test samples is available for development, the selection of the threshold can be done using that data. Yet, in many cases, no well-matched development data is available. In these cases, a threshold selected on some potentially mismatched development set would have to be used. Unfortunately, for modern SV systems, a threshold selected on a certain dataset is unlikely to be suitable for a different dataset, even when conditions are seemingly very similar~\citep{nandwana2019is}. 

As a consequence of this difficulty in the choice of threshold, SV systems can very rarely be used out-of-the-box. While automatic speech recognition (ASR) systems may be sub-optimal on conditions for which they have not been trained, in most cases they still return some reasonable output. On the other hand, SV systems with a threshold selected on conditions mismatched to the test conditions may make, on average, worse-than-random choices. Interestingly, in most cases, reasonably good performance can be achieved with the same system if the decision threshold is selected on the test data. That is, SV systems tend to have  reasonable discrimination performance even on conditions unseen during training but, since the score distributions are shifted and scaled depending on the sample's conditions, hard decisions on unseen conditions may be extremely poor, to the point of making the system useless.

The problem of threshold selection is highly related to the problem of calibration. In much of machine learning, a well-calibrated binary classifier is understood to map its input, called `data' below, to the class posterior:
$$
\text{data} \mapsto P(\text{class 1}\mid\text{data}) = 1-P(\text{class 2}\mid\text{data})
$$
In speaker verification, where class priors can vary widely between training and a variety of different applications at run-time, the \emph{log-likelihood-ratio} (LLR) format of calibration is preferred. The speaker verifier maps its input to the LLR as defined below: 
\begin{align}
\label{eq:LLRdef}
\text{data} \mapsto \text{LLR}=\log \frac{P(\text{data}\mid H_s)}{P(\text{data}\mid H_d)}
\end{align}
where `data' refers to all of the speech inputs for a speaker verification trial (or some processed version thereof), while $H_s$ is the \emph{same-speaker} hypothesis and $H_d$ is the \emph{different-speakers} hypothesis. The output of a well-calibrated classifier can be used to make minimum-expected-cost Bayes decisions. For the LLR format, the Bayes decision can be written as:
\begin{align}
\label{eq:BD}
\begin{split}
&\argmin_{d\in\Dset} \;\;P(H_s\mid\text{data})\ C(d\mid H_s)
+ P(H_d\mid\text{data})\ C(d\mid H_d) \\
=&\argmin_{d\in\Dset} \;\;\sigma(\text{LLR}+\gamma)\ C(d\mid H_s)
+ \sigma(-\text{LLR}-\gamma)\ C(d\mid H_d) \\
\end{split}
\end{align}
where $\sigma(x)=\frac1{1+e^{-x}}$ is the sigmoid function, $\gamma=\log\frac{P(H_s)}{P(H_d)}$ is the prior log odds, and $C(d\mid h)$ is the cost of making decision $d$, when $h$ is the true class. Typically, one takes $\Dset=\{\text{accept},\text{reject}\}$, where `accept' means deciding in favor of $H_s$ and `reject' deciding in favor of $H_d$. If correct decisions have zero cost:  $C(\text{accept}|H_s)=C(\text{reject}|H_d)=0$, then equation~\eqref{eq:BD} reduces to comparing the LLR to a threshold $t$ that is a function of $\gamma$ and $C$~\citep[see, e.g., ][]{Brummer:csl06}:
\begin{equation}
    t = \log \frac{C(\text{accept}|H_d)}{C(\text{reject}|H_s)} - \gamma. \label{eq:bayesthr}
\end{equation}

In summary, for a well-calibrated system where equation~\eqref{eq:LLRdef} holds for the test conditions, optimal decisions can be made using a threshold that solely depends on the cost function and the hypothesis prior. Hence, if we could develop a system that achieves robust calibration across conditions, we would, at the same time, solve the problem of threshold selection on unseen data. Also, for the same price, we would obtain scores that are interpretable as LLRs, which is essential in certain applications like forensic voice comparison. 

\begin{figure}[!t]
\centering
\includegraphics[width=1.02\columnwidth]{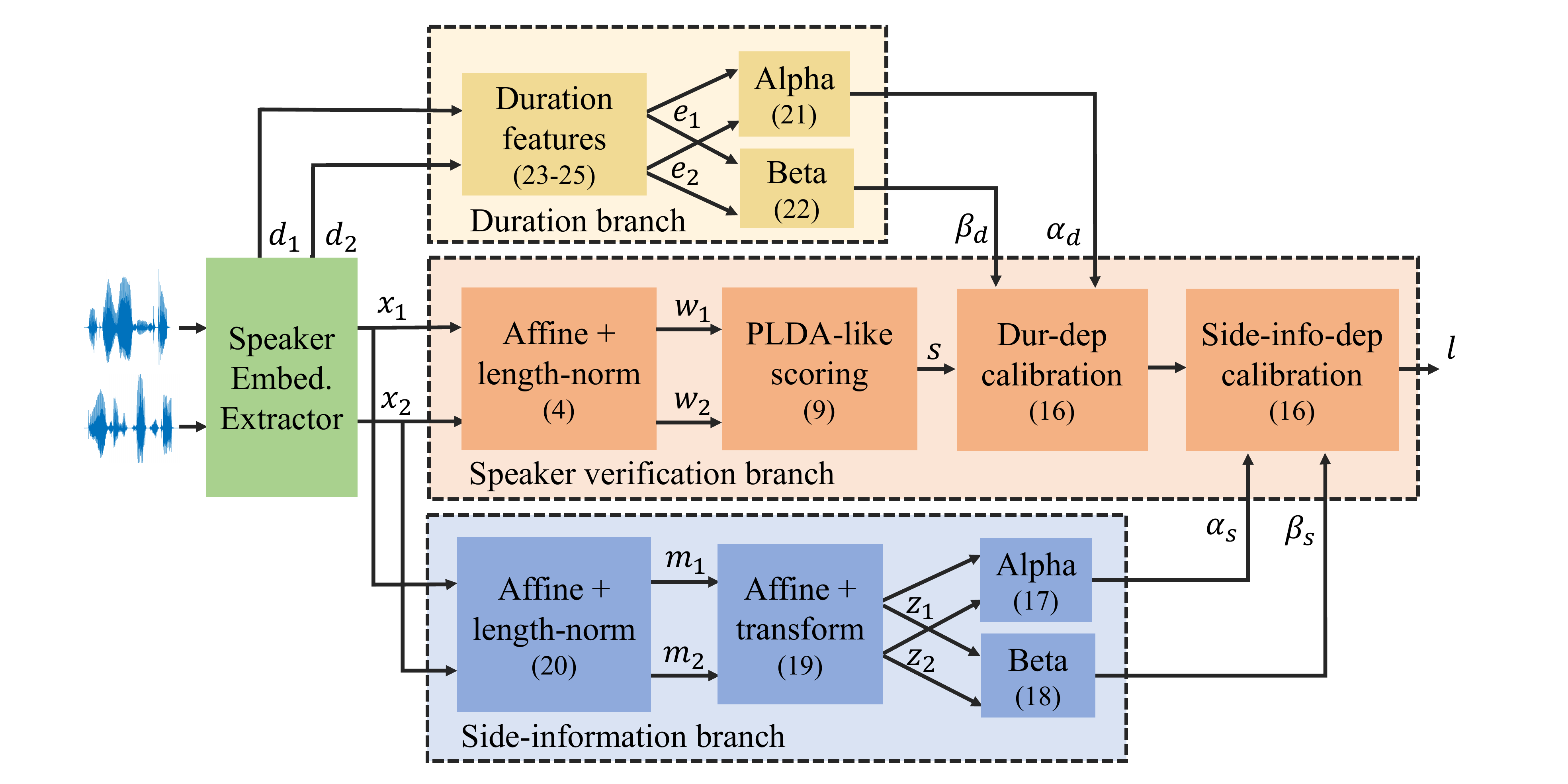}
\vspace{-0.6cm}
\caption{Schematic of the proposed DCA-PLDA backend. The backend includes the same components as the PLDA approach, an affine transform, length-normalization, PLDA-like scoring, and calibration, with the only difference in functional form being that the calibration stage, rather than being a global model with fixed parameters for all samples, is made to depend on duration features and side-information vectors extracted from the embeddings.
The equations indicated in parenthesis inside each block describe the functional form of the blocks and are explained in Sections \ref{sec:plda} and \ref{sec:dplda}.}  
\label{fig:architecture}
\end{figure}

In this paper we propose an SV system that  achieves excellent out-of-the-box calibration performance across a wide variety of conditions and, hence, does not require matched development data for the selection of the threshold for every new test set. As a side-effect, the system also improves discrimination on some datasets for which miscalibration is occurring within the set because they are composed of different conditions. The proposed approach modifies the standard \emph{probabilistic linear discriminant analysis} (PLDA) backend~\citep{kenny:10}, by adding condition-dependent calibration components. The initial idea was proposed in our prior work \citep{ferrer2020discriminative,ferrer2020speaker}. In this paper, we propose several improvements to the original method, including adding a duration-dependent calibration stage and an improved training procedure. A scheme of the architecture proposed in this paper can be found in Figure \ref{fig:architecture}.

We provide detailed analysis of the performance of the proposed approach for text-independent speaker verification, including new datasets not considered in our previous work, comparing different architectures and training data subsets, and showing the convergence behavior of our models.  To our knowledge, no other system in the literature has been shown to give a comparable level of robustness in terms of actual speaker verification performance across a wide variety of conditions. 

The code used to run the experiments in this paper and some example runs can be found in \url{https://github.com/luferrer/DCA-PLDA}.

\section{Prior Work}
\label{sec:prior_work}
Most current speaker verification systems are composed of a cascade of multiple stages. First, frame-level, acoustic features that represent the short-time contents of the signal are extracted, typically one feature vector every 10ms. These variable-length sequences of frame-level features are input to a deep neural network (DNN) which is trained to optimize speaker classification performance on the training dataset. The DNN uses a temporal pooling layer so that it can represent the speaker information in the variable-length input as a new feature of fixed dimension (typically 512) that is termed the \emph{speaker embedding} \citep[e.g.,][]{snyder2016deep}. The speaker embeddings are harvested from a hidden layer in the DNN, after the temporal pooling, but before the classifier. The speaker embeddings are then typically transformed using linear discriminant analysis (LDA), then mean- and, sometimes, variance-normalized and, finally, L2 length-normalized. Next, probabilistic linear discriminant analysis (PLDA) is used to obtain scores for each speaker verification trial. Finally, if required, a calibration stage can be included to convert the scores produced by PLDA into LLRs that can be used to make cost-effective Bayes decisions in the intended application. 

Although many different calibration strategies have been explored in the literature~\citep[e.g.,][]{brummer2014comparison}, a common solution is a simple, discriminatively trained affine transformation from scores to LLRs. The parameters of this transformation (a scale and an offset) are trained to minimize a weighted binary cross-entropy objective which measures the ability of the calibrated scores to make cost-effective Bayes decisions, when they are interpreted as LLRs~\citep{Brummer:csl06}. Assuming the calibration training data reflects the evaluation conditions, this procedure has been repeatedly shown to provide hard-to-beat performance on a wide range of datasets. Yet, when evaluation conditions differ from those present in the calibration training data, the average performance of the hard decisions made with the system can be very poor, sometimes even worse than that of random decisions.

Several approaches have been proposed in the speaker verification literature which take into account the signal's conditions in different ways in the calibration stage in order to achieve robustness across conditions. In some cases, the side-information --- some representation of the conditions of a signal --- was assumed to be known or estimated separately during testing \citep{FerrerEtAl:Eurospeech2005,Solewicz:Eurospeech2005,solewicz:ASLP07,FerrerEtAl:icassp2008,mandasari2013quality,mandasari2015quality,nautsch2016robustness}
 and calibration parameters were conditioned on these discrete or continuous side-information values. The Focal Bilinear toolkit \citep{FocalBilinear} implements a version of side-information-dependent calibration where the calibrated score is a bilinear function of the scores and the side-information vector, which is assumed to be composed of numbers between 0 and 1. More recently, we proposed an approach called trial-based calibration (TBC) where calibration parameters are trained independently for each trial using a subset of the development data \citep{mclaren2014trial,Ferrer:aslp18} selected using a model trained to estimate the similarity between the conditions of two samples. This approach, while successful, is computationally expensive. Another approach is proposed by \citet{tan2017vector} where the calibration stage jointly learns to calibrate and to estimate the SNR level and duration of the training signals, implicitly using the SNR and duration information to determine the calibration parameters. 

Our proposed approach follows the spirit of these earlier methods, tackling the problem of robustness in calibration by using a representation of the signals' conditions to determine the calibration parameters for each trial. The main differences between our method and the previous methods are (1) that the representation of the signal's conditions is estimated automatically within the model without the need for data labeled with respect to condition, in training or testing, and (2) that the full backend, including the score generation and the calibration components, is learned jointly and discriminatively to optimize speaker verification performance. The approach, first proposed in \citep{ferrer2020discriminative,ferrer2020speaker} and extended here, is a backend that takes speaker embeddings, for example, x-vectors \citep{snyder2016deep}, as input and outputs calibrated scores for each trial. The backend has the same functional form as the usual PLDA backend, including the calibration stage, but it is trained discriminatively to optimize a weighted binary cross-entropy. Further, the calibration stage is made to depend on the duration of the speech within the signal and on a low-dimensional side-information vector that is derived from the same speaker embeddings and is meant to represent the condition of the signal other than the duration. We call this method, discriminative condition-aware PLDA (DCA-PLDA).

The proposed method is related to recent papers~\citep{snyder2016deep,Rohdin2018,magnet:odyssey2020} which also propose to use the binary cross-entropy as loss function during DNN training. 
The idea of discriminatively training a PLDA-like backend was first proposed by \citet{burget:icassp11}. In that work, the parameters of the backend were trained using a support vector machine or linear logistic regression. More recently,~\citet{snyder2016deep} applied this idea to a DNN-based SV pipeline, integrating the embedding extractor and the backend together and training them jointly to optimize binary cross-entropy. A similar approach was presented by \citet{ramoji2020neural}, who proposed to use a loss function that approximates the detection cost function instead of binary cross-entropy.
\citet{Rohdin2018} proposed to use an architecture that mimics the previous i-vector pipeline~\citep{Dehak11}  for speaker verification, pre-training all parameters separately and then fine-tuning the full model to minimize binary cross-entropy. 
Finally,~\citet{magnet:odyssey2020} proposed to transform the scores produced as the dot product between the embeddings generated by a ResNet with an affine function trained to optimize binary cross-entropy. The weight parameter of this transform is given by a network that takes as input the pooling layer from the ResNet. 

Most of the papers described above focus on discrimination performance, not including calibration performance in the results. 
One paper that shows calibration performance is the one by \citet{magnet:odyssey2020}, where both discrimination and calibration performance are shown to improve using the proposed method over a global calibration model. Yet, the calibration loss for some conditions is still very large. 
Our proposed method is similar to this method, with the main difference being that, in our case, the calibration
stage depends on the speech duration, as well as a low-dimensional estimate of the condition of each sample.
The dependency on the speech duration, according to our results, appears to be key for giving good calibration generalization.
Yet, several other characteristics, including training and test data, network architecture, and training procedure differ in both methods which might also explain the difference in calibration performance.    

Finally, another family of approaches that are related to the proposed method are those that perform adaptive normalization \citep{karam2011towards,cumani2011comparison,matejka2017analysis}. In these approaches, the score obtained for a certain trial is normalized using the statistics obtained by scoring the samples in the trial against a subset of samples selected from a \emph{cohort} set. The subset of cohort samples used for computing the statistic for each trial is selected based on their similarity to the trial's samples. These methods, hence, have the potential to be able to mitigate miscalibration problems by aligning the scores from different conditions to each other. Yet, note that the normalized scores are not calibrated. A final calibration step would be needed to convert the normalized scores into likelihood ratios. 

\section{Standard PLDA-based Backend}
\label{sec:plda}
In the current state of the art in embedding-based speaker verification, two scoring approaches are common. One is cosine scoring, where the score is the dot product between two L2 length-normalized embeddings~\citep[e.g.,][]{magnet:odyssey2020}. This approach has no trainable parameters in the scoring stage and instead relies on training the embedding extractor in a way that works well with cosine scoring. Although cosine scoring can achieve very good accuracy on matched test and training conditions, it is not equipped to deal with variable conditions and there is no easy way to adapt it to new conditions. The other way to score is to use a trainable PLDA-based backend. PLDA training requires less data than what is needed for training an embedding extractor and the computational requirements are orders of magnitude smaller. PLDA scoring is therefore an attractive way to adapt a speaker recognizer to new conditions. PLDA backends are typically composed of several stages, described in the following sections.

\subsection{Pre-processing}
First, linear discriminant analysis (LDA) is applied to reduce the dimension of the embeddings while emphasizing speaker information and reducing other irrelevant information. In addition to concentrating speaker information, the LDA stage also contributes to making the data more Gaussian: every output of the LDA projection is a sum of many relatively independent inputs, which tend to be more Gaussian. Then, each transformed dimension is mean- and variance-normalized and the resulting vectors are length normalized, a procedure that has an even more powerful Gaussianization effect, which is essential for good performance of the subsequent Gaussian PLDA modeling stage~\citep{romero:lennorm}. The pre-processing stage for an embedding $x$ of dimension $D$ can be summarized by the following equation:
\begin{eqnarray}
w \hspace{-0.2cm} & = & \hspace{-0.2cm}\Norm( A_p x+ m_p),\label{eq:lda} 
\end{eqnarray}
where $A_p$ is a matrix of size $N\times D$ containing the first N LDA directions. The rows in this matrix are scaled so that the resulting vectors $w$ for the training data have a variance of 1.0 in each dimension. The vector $m_p$, of dimension $N$, is the global mean of the data after multiplication with $A_p$. $\Norm$ performs L2 length normalization. 

\subsection{PLDA}
\label{sec:plda_plda}
After pre-processing, PLDA \citep{Ioffe:2006,prince:plda}  is used to compute a score for each trial. The two-covariance PLDA variant~\citep{Brummer:odyssey10}, widely used in speaker verification, assumes that each pre-processed embedding, $w$, can be modeled as
\begin{equation*}
w = y + e,
\end{equation*}
where $y$ and $e$ are vectors of dimension $N$, are assumed to be independent, and are both Gaussian-distributed so that
\begin{eqnarray*}
y & \sim & \mathcal{N}(\mu,B^{-1}), \\
w|y & \sim & \mathcal{N}(y, W^{-1}),
\end{eqnarray*}
where $B$ is the between-speaker precision matrix and $W$ is the within-speaker precision matrix, both with dimension $N\times N$. 

The training procedure used to obtain point estimates for $\mu$, $B$ and $W$ requires the use of an expectation-maximization algorithm  \citep{em4splda,sizov2014}. Initial values $\mu_0$, $B_0$ and $W_0$ for the parameters are needed in order to jump start the EM algorithm. While random initialization works fine, a much better starting point can be achieved by using the sample estimates of the global mean and covariance matrices. 
Given a set of pre-processed vectors $w_i$ where sample $i$ corresponds to speaker $s_i$, we compute the initial parameters as follows:
\begin{eqnarray}
\mu_s & = & \frac{\sum_{i|s_i=s} w_i}{N_s} \nonumber \\
\mu_0 & = & \frac{\sum_{s} c_s \sum_{i|s_i=s} w_i}{\sum_{s} c_s N_s} \label{eq:mu0}\\
B_0 &=& \left[ \frac{\sum_s  c_s N_s  ( \mu_s - \mu_0) (\mu_s -\mu_0)^\T}{\sum_{s} c_s N_s}\right]^{-1} \label{eq:B0}\\
W_0 &=& \left[ \frac{\sum_{s} c_s \sum_{i|s_i=s}  ( x_i - \mu_s) (x_i - \mu_s)^\T}{\sum_{s} c_s N_s} \right]^{-1} \label{eq:W0}
\end{eqnarray}
where $i$ is an index over samples, $s$ is an index over speakers, $N_s$ is the number of samples for speaker $s$, and $c_s$ is an arbitrary weight corresponding to speaker $s$ which can be used to increase or decrease the influence of a speaker's data. The sums over $i|s_i=s$ operate over all the samples corresponding to speaker $s$. As we will see, in our experiments, the weights are used to compensate for the severe imbalance across conditions that is present in our training data. 
Note that, for integer weights, this weighting procedure is equivalent to repeating the data from each speaker $s$ $c_s$ times, taking each repetition as coming from a different speaker. 

The EM formulation can also be changed easily to introduce the weights in the expectation step. Specifically, the weights are introduced in equations (24) and (25) in \citep{em4splda} by multiplying each term in each of those summations with the weight corresponding to speaker $i$ (note that subindex $i$ in those equations is the speaker index).

Given a trial composed of a single enrollment and single test sample with pre-processed embeddings, $w_1$ and $w_2$, the PLDA score is defined as the following log-likelihood-ratio (LLR):
\begin{eqnarray}
s & = & \log \frac{P(w_1, w_2|H_s)}{P(w_1, w_2|H_d)}
\end{eqnarray}
where, as mentioned in Section \ref{sec:intro}, $H_s$ and $H_d$ are the hypotheses that the enrollment and test speakers are the same or different, respectively.
The following closed-form solution can be derived for this LLR given the assumptions for the PLDA model:
\begin{eqnarray}
s  \hspace{-0.2cm} & = & \hspace{-0.2cm}2 w_1 ^\T \Lambda w_2 + w_1^\T \Gamma w_1 + w_2^\T \Gamma w_2 + w_1^\T c + w_2^\T c + k,  \label{eq:plda_score}
\end{eqnarray}
where $\Lambda$ and $\Gamma$ are matrices of size $N\times N$, $c$ is a vector of size $N$, and $k$ is a scalar.
These parameters can be obtained as a function of $B$, $W$ and $\mu$:
\begin{eqnarray}
\Lambda & = & \frac{1}{2} W^\T \tilde \Lambda W \\
\Gamma & =  & \frac{1}{2} W^\T (\tilde \Lambda - \tilde \Gamma) W \\
c & = & W^\T (\tilde \Lambda - \tilde \Gamma) B \mu \\
k & = & \frac{1}{2} \tilde k + \frac{1}{2} (B \mu)^\T (\tilde \Lambda - 2 \tilde \Gamma) B \mu \label{eq:k}
\end{eqnarray}
with 
\begin{eqnarray*}
\tilde \Lambda & = & (B+2W)^{-1} \\
\tilde \Gamma & = & (B+W)^{-1}\\
\tilde k & = & - 2 \log |\tilde \Gamma| - \log |B| + \log |\tilde \Lambda| + \mu^\T B \mu
\end{eqnarray*}

See \citep{cumani2013pairwise} for a derivation of these equations. Note, though, that the expression for $k$ in that paper is missing a 1/2 multiplying $\tilde k$ and $\tilde k$ has wrong signs for two of the terms. These errors have been corrected in the equations above.

In summary, once the PLDA parameters have been trained, the score for each trial can be computed with a simple second-order polynomial expression of the pre-processed embeddings for the trial.

\subsection{Calibration}
\label{sec:global_calibration}

As mentioned above, PLDA scores are computed as LLRs given the estimated PLDA model's parameters. Yet, since the assumptions made by PLDA do not exactly hold in practice, the scores produced by this model are usually badly calibrated.  For this reason, the usual procedure is to post-process the PLDA scores using a calibration stage.
The standard procedure for calibration in speaker verification is to use linear logistic regression, which applies an affine transformation to the scores, training the parameters to minimize binary cross-entropy \citep{brummer2013likelihood}.  The objective function to be minimized is given by
\begin{equation}
C_\pi = -\frac{\pi}{T} \sum_{i \in \cal T}  \log(q_i) - \frac{1-\pi}{N} \sum_{i \in \cal N}  \log(1-q_i), \label{eq:crossent}
\end{equation}
where
\vspace{-0.2cm}
\begin{eqnarray}
q_i & = &\sigma\left(l_i + \gamma \right), \\
l_i & = & \alpha s_i + \beta, \label{eq:cal}
\end{eqnarray}
and where $s_i$ is the score for  trial $i$ given by equation (\ref{eq:plda_score}), $\sigma$ is the sigmoid function, $\pi=P(H_s)$ is a parameter reflecting the expected prior probability for same-speaker trials, $\gamma = \log (\pi /(1-\pi))$ is the prior log odds, and $\alpha$ and $\beta$ are scalar calibration parameters to be optimized by minimizing the quantity in equation~(\ref{eq:crossent}).

Since the cross-entropy is calibration-sensitive~\citep{brummer2013likelihood}, this procedure usually leads to well-calibrated scores on data that has class-conditional score distributions that are similar to those of the data used to train the calibration model. On the other hand, this does not guarantee that calibration will be good on data from any other condition, where score distributions may differ. \citet{nandwana2019is} showed the effect that a mismatch in several different characteristics of the signals has on calibration performance. In particular, they show that speech duration, distance to the microphone and language mismatch can cause a large degradation in calibration performance.  Those results illustrate a very common phenomenon in speaker verification: applying a calibration model on data mismatched to that used to train the model can lead to severely miscalibrated scores. This is the problem we aim to tackle in this work.

\section{Theoretical perspective on conditions}
\label{sec:condition}
\def\xvec{w}
\def\zvec{y}
\def\Gmat{G}
\def\Fmat{F}
Thus far we have relied on an intuitive understanding of the concept of the \emph{conditions} of speaker verification trials. In this section we present a generative modeling view that explains in more detail what we mean by conditions and how they interact with the PLDA model. We motivate why the PLDA model might be too simple, and why the more complex model that is proposed in this paper could perform better.

Consider the following graphical model for a speaker verification trial, where a pair of pre-processed speaker embeddings, $\xvec_1$ and $\xvec_2$ have been observed and where we need to infer the unknown hypothesis, $h\in\{H_s,H_d\}$:
\begin{align*}
\begin{tikzpicture}
\node[obs] (x1) {$\xvec_1$};
\node[latent, right = of x1] (z1) {$\zvec_1$};
\node[latent, left = of x1] (e1) {$e_1$};
\node[latent, right = 5em of z1] (z2) {$\zvec_2$};
\node[obs, right = of z2] (x2) {$\xvec_2$};
\node[latent, right = of x2] (e2) {$e_2$};
\node[const, right = of e2] (plda) {PLDA};
\node[latent, below = 5em of e1] (c1) {$\eta_1$};
\node[obs] (d1) at(c1-|x1) {$d_1$};
\node[latent] (g1) at(c1-|z1) {$g_1$};
\node[latent] (g2) at(c1-|z2) {$g_2$};
\node[latent] (c2) at(c1-|e2) {$\eta_2$};
\node[obs] (d2) at(c1-|x2) {$d_2$};
\node[latent, below right = 3em of z1] (h) {$h$};
\node[const] (cond) at(c2-|plda) {conditions};
\edge {z1,e1,g1}{x1};
\edge {z2,e2,g2}{x2};
\edge {c1}{x1};
\edge {c2}{x2};
\edge {g1}{g2};
\edge {z1}{z2};
\edge {d1}{x1};
\edge {d2}{x2};
\edge{h}{z2,g2}
\end{tikzpicture}
\end{align*}
Observed variables are shaded and latent ones are clear. Under the same-speaker hypothesis, $h=H_s$, the variables associated with the speaker are tied: $\zvec_2=\zvec_1$ and $g_2=g_1$. Otherwise, when $h=H_d$, then $\zvec_2,g_2$ are generated independently from $\zvec_1,g_1$. The top row represents the PLDA model, while in the bottom row, we have introduced a collection of variables, which we refer to as the \emph{conditions} of the trial. 

Before the embeddings are observed, $h$ is dependent\footnote{See for example~\cite{bishop:ml} for an explanation of how to read conditional independence relationships from graphical diagrams.} only on $\zvec_i,g_i$, and \emph{not} on $e_i,\eta_i,d_i$. The former group therefore carries speaker information, while the latter does not. The condition variables include \emph{both} types. After the $\xvec_i$ have been observed, $h$ depends on all other variables in the graph, which shows that even the variables that do not carry speaker information by themselves, can help to better infer $h$.  

In PLDA, $\zvec_1,\zvec_2$ are the \emph{hidden speaker identity variables}, theoretical variables suggested by Patrick~\cite{Kenny:JFA} that determine speakers' voices. The hidden variables, $e_1,e_2$ represent additive Gaussian noise that model the variability between recordings of the same speaker. PLDA training and scoring algorithms marginalize over these hidden variables \citep[see, e.g.,][]{kenny:10,Brummer:odyssey10}. 

The condition variable $d_i$ is the duration of the speech within the segment from which the embedding, $\xvec_i$ has been extracted. We will refer to $d_i$ as the \emph{duration}, for short. In this paper, we take the $d_i$ as observed and involve them explicitly in our proposed calculations. In practice, $d_i$ will usually be given by a speech activity detection system and, hence, the performance of the approach will depend on the performance of that system. The other condition variables, $\eta_i,g_i$ are hidden and we do not identify them explicitly. We do not need labels for these variables in the data and we do not explicitly reference these variables in our algorithms. Nevertheless, it is useful to suppose that there are such variables, that enrich the purely Gaussian PLDA model to a more complex one that would better represent real-world data.

We can think of the $g_i$ as a collection of factors that roughly determine the speaker's vocal qualities---for example gender and mother tongue. The $\eta_i$ do not carry speaker information, but include things like recording device (laptop, telephone, etc.), spoken language, noise level, microphone placement, sampling rate, speaking style and so on. Among these factors, the discrete ones could generalize PLDA to a mixture model, while continuous factors could scale the Gaussian PLDA variables, resulting in heavy-tailed PLDA, as explained in~\citep{kenny:2010, Silnova_HT}.

The classical PLDA training and scoring algorithms effectively marginalize over the $y_i$ and $e_i$ variables at \emph{fixed}, implicit values of the condition variables. The work in this paper effectively augments PLDA with mechanisms that can deal with \emph{variable} conditions. These mechanisms, which will be detailed in the next section, can be understood in two different ways:  
\begin{itemize}[leftmargin=1.5em]\itemsep0em
    \item Explicitly, we are discriminatively training a data-dependent calibrator jointly with the PLDA model.
    \item Implicitly, we are replacing the traditional combination of a PLDA model and a fixed post-calibrator, with a more complex model that produces well-calibrated scores across conditions. The above graphical diagrams represent the new model. Since this model is discriminatively trained without any explicit reference to the hidden variables, it is implicitly learning a scoring function where all the hidden variables (the PLDA ones and the new condition variables) are effectively marginalized.         
\end{itemize}

\section{Condition-Aware Discriminative Backend}
\label{sec:dplda}

In our recent papers \citep{ferrer2020discriminative,ferrer2020speaker} we presented a backend method with the same functional form as the PLDA-backend explained in Section \ref{sec:plda}, but where all parameters are optimized jointly, in a manner similar to the one used by \citet{Rohdin2018} (though, note that in this paper we only optimize jointly the backend stage instead of the full pipeline, as in Rohdin's paper). The novel aspect in our previous papers was the integration into the backend of a side-information-dependent calibration stage aimed at improving  calibration robustness across conditions. We proposed to capture the information about the conditions that affect calibration in a side-information vector, which was then used to determine the parameters of the calibration stage. In our first paper  \citep{ferrer2020discriminative}, the side-information vector was generated by a separate model, which was trained to predict labeled conditions in the training data. These condition labels were derived from the available metadata in each of the training datasets, which had different levels of detail depending on the dataset. In our second paper, we integrated the extraction of the side-information vector within the backend itself. This approach has the advantage of not requiring condition-labeled data during training. Further, it gives the backend the freedom to define as side-information only what is useful for calibration. In this paper, we continue in the direction of the second approach, with the addition of a duration-dependent calibration stage. Hence, in the current paper the condition of a sample is represented by both the side-information vector as well as the duration of the speech in the sample. 
Note that while the speech duration can also be considered a type of side-information, as we will see, the addition of a dedicated duration-dependent branch
leads to a substantial gain over the original architecture. In the following subsections, we describe the backend architecture and the training procedure in detail.

\subsection{Formulation}
\label{sec:formulation}

In Section \ref{sec:results} we show that discriminatively training a PLDA-like backend does not suffice to obtain good calibration across conditions. This is, in part, due to the fact that the PLDA assumptions (explained in Sections \ref{sec:plda_plda} and \ref{sec:condition}) do not hold in practice. This problem can be solved, as usually done for the standard PLDA backend, by training a specific calibration model for each domain of interest, which requires having at least some domain-specific labeled data. 
In this research, though, we assume that no domain-specific data is available for system adaptation or for training a calibration model. This also means that a domain-specific decision threshold cannot be learned. Hence, we aim to design the best possible out-of-the-box system for unknown conditions for which the output scores will be well-calibrated even on such unseen conditions. As explained in Section \ref{sec:intro}, well-calibrated scores, when interpreted as LLRs, can be thresholded at theoretically-determined Bayes decision thresholds, which will then make cost-effective decisions for a range of different applications \citep[see, for example,][]{van2007introduction}. 

In order to achieve the goal of a robust well-calibrated system, we make the parameters of the calibration stage depend on side-information vectors that are meant to describe the conditions of the signals that affect calibration. As explained in Section \ref{sec:condition}, the goal behind this approach is to generalize the PLDA formulation to take into account the condition of the signals. Specifically, we transform the scores produced by PLDA using an affine transformation as in equation \ref{eq:cal}, but make the parameters $\alpha$ and $\beta$ be functions of side-information vectors, $z_1$ and $z_2$, for each of the signals in a trial, using a form identical to the one used for PLDA scoring:
\begin{eqnarray}
\alpha_s  &=&  2 z_1^\T \Lambda_{\alpha_s} z_2 + z_1^\T \Gamma_{\alpha_s} z_1+ z_2^\T \Gamma_{\alpha_s} z_2 + (z_1 + z_2)^\T c_{\alpha_s} + k_{\alpha_s}, \label{eq:alpha_s} \\
\beta_s  &=&  2 z_1^\T \Lambda_{\beta_s} z_2 + z_1^\T \Gamma_{\beta_s} z_1 + z_2^\T \Gamma_{\beta_s} z_2 + (z_1 + z_2)^\T c_{\beta_s} + k_{\beta_s}. \label{eq:beta_s} 
\end{eqnarray}
where the $\Gamma$'s, $\Lambda$'s, $c$'s, and $k$'s are parameters to be optimized. The key component of this model are the side-information vectors,  $z_i$, which we define to be given by
\begin{equation}
z_i = f(A_z m_i + b_z), \label{eq:z}
\end{equation}
where $A_z$ and $b_z$ are parameters to optimize, $f$ is a transformation which is empirically selected from a small set of options described below, and $m_i$ is obtained using the same form as in equation (\ref{eq:lda}), though with different parameters, $A_m$ and $b_m$. That is, 
\begin{equation}
m_i = \Norm(A_m x_i + b_m), \label{eq:m}.
\end{equation}
The dimensions of $m_i$ and $z_i$ are hyperparameters of the model, which are optimized empirically.
We explored three different forms for function $f$: the identity function where no transformation is used, softmax, and the logarithm of the softmax, which is what we used in our prior work \citep{ferrer2020discriminative,ferrer2020speaker}. 
During development we tried several other architectures for the side-information branch, including feed forward neural networks which concatenated the embeddings from the two sides of the trial at the input or after a layer of separate processing, generating $\alpha_s$ and $\beta_s$ at their output. None of these alternative approaches outperform the one described above, possibly due to increased complexity of the models which more easily resulted in overfitting.

Finally, in this work we propose to add an additional duration-dependent calibration stage before the side-information-dependent one. Several works have shown that sample duration has a large impact on the score distribution \citep[e.g.,][]{nandwana2019is,mandasari2013quality}. As a consequence of this, training a calibration model with data of durations that are mismatched to the test data can lead to severely miscalibrated scores. Different approaches have been proposed in the literature to mitigate this problem \citep{mandasari2013quality,mandasari2015quality,nautsch2016robustness}. In these works, the duration of each of the samples in a trial is used to condition the parameters of the calibration model. Here, we propose a simple model where the speech duration of the samples in a trial are transformed into feature vectors and the calibration parameters are obtained as a function of these vectors using the same functional form as in equations (\ref{eq:alpha_s}) and (\ref{eq:beta_s}). 
\begin{eqnarray}
\alpha_d & =&  2 e_1^\T \Lambda_{\alpha_d} e_2 + e_1^\T \Gamma_{\alpha_d} e_1+ e_2^\T \Gamma_{\alpha_d} e_2 + (e_1 + e_2)^\T c_{\alpha_d} + k_{\alpha_d}, \label{eq:alpha_d} \\
\beta_d   & = & 2 e_1^\T \Lambda_{\beta_d} e_2 + e_1^\T \Gamma_{\beta_d} e_1 + e_2^\T \Gamma_{\beta_d} e_2 + (e_1 + e_2)^\T c_{\beta_d} + k_{\beta_d}. \label{eq:beta_d} 
\end{eqnarray}
where the $\Gamma$'s, $\Lambda$'s, $c$'s, and $k$'s are parameters to be optimized, and $e_i$ is a feature vector derived from $d_i$, the number of frames used to extract the embedding vector $x_i$. 

In this work we explore three ways to convert a duration $d$ into a feature vector, given by
\begin{eqnarray}
e_{\text{log}}(d) & = & \log(d), \label{eq:el} \\
e_{\text{bin}}(d) & = & \onehot(\bin(d, t)), \label{eq:eb} \\
e_{\text{wlog}}(d) & = & \log(d)\ [\sigma_{sc}(d), (1-\sigma_{sc}(d))]^\T, \label{eq:elb}\\
\sigma_{sc}(d) & = & \sigma\left(s\ (\log(d)-\log(c))\right) \nonumber
\end{eqnarray}
where $\bin$ refers to the operation that turns the $d$ into an index corresponding to its bin given a list of thresholds $t$, $\onehot$ converts the index into a vector with a 1 at that index and 0 elsewhere, $\sigma$ is the sigmoid function, and $s$ and $c$ are hyperparameters to be tuned. Note that $e_{\text{log}}$ is a scalar, $e_{\text{bin}}$ is a column vector of size given by the number of thresholds in $t$ plus 1, and $e_{\text{wlog}}$ is a column vector of size 2. 

In the first case, $e_{\text{log}}$, we assume that the calibration parameters can be estimated as a second-order polynomial function of the log duration of the two samples involved in the trial.  
In the second case, the calibration parameters are free to take any possible value for each combination of enrollment and test bins. This gives more flexibility to the model, but has the disadvantage of setting the calibration parameters to fixed values within each combination of enrollment and test bins. Hence, in this approach it is necessary to use several bins so that the piece-wise constant approximation can fit the data. This may potentially result in a model that overfits the training conditions more easily. Note that in this case, one does not need to use the full expression for $\alpha_d$ and $\beta_d$; the $\Lambda$ term suffices since the one-hot vectors operate as selectors of the $\alpha_d$ and $\beta_d$ for each combination of bins for the two samples in the trial. 
Finally, the third option, $e_{\text{wlog}}$ is a hybrid where we assume that the second order polynomial of the log durations is approximately appropriate over the left and right regions of the duration space, defined by the parameter $c$. By multiplying the logarithm of the duration with a sigmoid centered at $\log(c)$ and its flipped version, we create two features that are non-zero for small and large durations, respectively, with tapering at values near $\log(c)$ determined by the $s$ parameter. 
This approach allows more flexibility than using the single logarithmic feature while still being continuous. We compare these three approaches in Section \ref{sec:results}.

Figure \ref{fig:architecture} on page~\pageref{fig:architecture} shows the complete architecture of the proposed backend. Note that the orange blocks at the center correspond to the standard PLDA formulation, except that the global calibration stage is replaced by duration and side-information-dependent calibration stages.

\subsection{Training Procedure}

The proposed architecture can be trained using stochastic gradient descent or one of its variants. In particular, in this work we use Adam optimization~\citep{kingma2014adam}. These algorithms require an initial value for the parameters, the selection of a loss function, and a process to generate batches during training. We discuss all these issues in this section. Further, we explain the training procedure, which is done in stages, and the way we select the best epoch from each run.

\subsubsection{Initialization}
\label{sec:init}

We initialize the parameters in the speaker verification branch (Figure \ref{fig:architecture}) using the corresponding values from a standard PLDA-based backend, which are obtained as described in Section \ref{sec:plda}. The parameters of the calibration stages are all initialized to 0, except for the $k$ values, which are initialized as follows:
\begin{itemize}[leftmargin=1.5em]\itemsep0em 
    \item If both calibration stages are used, then the $k$ parameters of the duration stage are initialized to the global calibration parameters, obtained using linear logistic regression as explained in Section \ref{sec:global_calibration}, and the ones for the side-information stage are initialized to be a pass-through, $k_{\alpha_s} = 1$ and $k_{\beta_s} = 0$.
    \item If only one stage is used (duration- or side-information-dependent), then the $k$ parameters for the $\alpha$ and $\beta$ blocks entering that model are initialized to the corresponding global calibration parameters.
    \item Finally, in some of our experiments, we test a plain discriminative PLDA (D-PLDA) architecture were the calibration stage has fixed values for $\alpha$ and $\beta$ that do not depend on duration or side-information. In this case, those parameters are initialized to the global calibration parameters. 
\end{itemize}
 
Given this initialization procedure, before the first training iteration the model is identical to a PLDA-based backend with global calibration. This gives us a way to directly compare PLDA, D-PLDA and DCA-PLDA systems using exactly the same code-base, which can be found in \url{https://github.com/luferrer/DCA-PLDA}.

The side-information branch has some additional parameters that need to be initialized. This initialization has no effect on the output of the model before the first training iteration, since, as described above, the $\Gamma$, $\Lambda$ and $c$ parameters in equations (\ref{eq:alpha_s}) and (\ref{eq:beta_s}) are initialized to 0. Yet, this initialization has a considerable effect on the performance of the final model. To initialize the first stage of the side-information branch, we use the last M dimensions of the full LDA transform (before selection of the best N dimensions) used to initialize the speaker verification branch of the model, including mean and variance normalization. That is, we use the dimensions that are less useful for speaker verification under the LDA assumptions and apply the same procedure as the LDA stage in the speaker verification branch. In \citep{ferrer2020speaker} we show that this initialization gives an advantage over random initialization.
The parameters of the dimensionality reduction stage (equation \ref{eq:z}) are the only parameters that are initialized randomly, using a normal distribution centered at 0.0 with standard deviation of 0.5. 

After initialization, the $\Gamma$ and $\Lambda$ parameters in all polynomial blocks are symmetric matrices. During training, these matrices are still constrained to be symmetric. This is done simply by optimizing auxiliary matrices $\hat \Gamma$ and  $\hat \Lambda$, and setting $\Gamma = 0.5\ (\hat \Gamma + \hat \Gamma^T)$ and $\Lambda = 0.5\ (\hat \Lambda + \hat \Lambda^T)$.

\subsubsection{Loss Function and Batch Generation}
\label{sec:batches}
We train the model for the target speaker verification task. To this end, we use the same loss function used to train the calibration model in the traditional systems, the weighted cross-entropy given in equation (\ref{eq:crossent}), and optimize it using the Adam optimizer. Hence, we need to define mini-batches over which the loss and gradients are computed at each optimization step. These mini-batches need to be composed of speaker verification trials. In this work, we restrict ourselves to solving the simplest speaker verification problem where two speech segments are compared to each other to decide whether the two belong to the same speaker or not. That is, each trial is simply composed of two audio segments. 

We test two different ways of creating the trials for a batch of size $N$, where $N$ is an even number. In the simplest method, similar to the one used in our previous papers, we select samples as follows:
\begin{itemize}[leftmargin=1.5em]\itemsep0em 
    \item First, select $N/2$ speakers.
    \item For each of those speakers, select two different recording sessions.\footnote{Two speech segments are defined as belonging to the same session if they were recorded during the same day, in the same recording setup; for example, when the same audio is recorded using more than one microphone, or when several recordings are made one after the other. Also, during a telephone conversation, both sides of the conversation are considered to belong to the same recording session. Finally, when we split waveforms in chunks or when we degrade them as described in \ref{sec:appendix_embeddings}, all chunks and degraded versions deriving from a certain waveform belong to the recording session of the original waveform.}
    \item For each selected session, select one sample.
\end{itemize}
Speaker, sessions and samples are selected, in order, from randomly sorted lists. That is, at the beginning of each training process, we randomly sort a set of lists of speakers, sessions per speaker and samples per session. Then, when creating mini-batches, we traverse those lists in order, selecting the next item in the list until the list is exhausted, at which time we randomize the list again and start selecting from the beginning of that list.  In this way, during training we see each speaker approximately the same number of times. For each speaker, each of their sessions is also seen approximately the same number of times. And, for each session, each of its samples is seen approximately the same number of times.  Finally, once the samples for a mini-batch have been selected, all possible trials between these N  samples are used to compute the cross-entropy, excluding different-domain and same-session different-speaker (impostor) trials. 

In the second method for batch selection, we modify the method above to ensure that each batch is composed of the same number of speakers from each training domain. That is, if the training data is composed of D domains, then at the first step during batch creation we select N/D/2 speakers from each domain. The rest of the process is identical to the first method.

\subsubsection{Training Stages and Model Selection}
\label{sec:train_stages}
In our previous works we trained the parameters using a two-stage procedure where the parameters of the speaker verification branch were trained using all available samples during the first stage and then frozen during a second stage where the parameters of the side-information branch and the calibration stage were fine-tuned using data balanced by domain. During both stages the mini-batches were created using a method similar to the first one described above without domain balancing. Yet, since during the second stage the input data was domain-balanced, the mini-batches created during that stage were also approximately domain-balanced. 

In the current work, instead of creating a balanced dataset, which results in a large percentage of the data being unused during training, we use the second method for batch generation described above, where the domain balance is enforced at batch creation. In this case, all training speakers are used. In Section \ref{sec:results} we compare two procedures for training: (1) balancing by domain at initialization and during batch generation, and (2) without balancing at either stage.

As mentioned above, we want the resulting model to be robust to a variety of conditions, including those that may not appear during training. What we find experimentally is that the models trained with the procedure above vary wildly  in terms of robustness from one mini-batch to the next. That is, while the loss computed on the training data or on data from conditions similar to those in training behaves as expected, decreasing somewhat monotonically as training progresses (though, certainly, with some noise), the loss on unseen conditions is extremely noisy, even from one mini-batch to the next. Results illustrating this phenomenon are shown in Section \ref{sec:learning_curves}. 

Given this behavior, we use the following three-stage procedure to train the models:
\begin{itemize}[leftmargin=1.5em]\itemsep0em 
    \item Stage 1: After initializing the parameters as described in Section \ref{sec:init}, we train the model over several thousand mini-batches to reach good performance on seen conditions. 
    \item Stage 2: The last model from the first stage is then further trained using a higher learning rate than in the first stage, evaluating the loss on a list of development sets after every mini-batch update. The increase in learning rate was found to give improved results over using the same rate as in the first stage. After training with the increased learning rate over a few thousand mini-batches, we select the model that led to the best average loss over the development sets. 
    \item Stage 3: The best model from the second stage is fine-tuned using a slower learning rate over a few more batches. The final model from the training procedure is given by the best model from the third stage, selected based on the average performance over the development sets. 
\end{itemize}

Finally, we found the random seed to have a non-negligible effect on the performance of the resulting model. The random seed affects the mini-batch creation and also the initialization of the dimensionality reduction stage that turns $m$ vectors to $z$ vectors. In our experiments, unless otherwise stated, the results shown correspond to the best model over 20 seeds, selected based on the average performance over the development sets.

\section{Experimental Setup}
In this section we describe the system configuration and datasets used for our experiments. 
We use standard x-vectors as input for our backend experiments \citep{snyder2016deep,KaldiRecipe17}. Details on the embedding extractor, including a description of the input features, speech activity detection, training data, and procedure can be found in \ref{sec:appendix_embeddings}.

\subsection{Backend Training Data}
\label{sec:backend}
The training data for the PLDA and DCA-PLDA backends includes all the training data used for the embedding extractor, plus two additional datasets. We divide this data in six different domains: VOX, SWB, MIX, MIX ML, FVCAUS and RATS SRC, where the first four domains are the ones used for embedding extractor training. Details on the source datasets and characteristics of each of these domains can be found in \ref{sec:appendix_backend_data}.
For backend training we explore two options; using the full waveforms available from each domain, or cutting each waveform into smaller segments from 4 to 240 seconds, creating 8 chunks per waveform. The chunks are created to achieve a close-to-uniform distribution of log durations for each domain. 

\subsection{Development and Evaluation Datasets}
\label{sec:evaldata}

We focus on the problem of text-independent speaker verification, with trials defined as the comparison between two individual segments each containing a single speaker. We use several different datasets for development and evaluation of the proposed approach to include a large variety of conditions, including six sets extracted from each of the six training domains, the SITW dataset, data from the speaker recognition evaluations (SRE) from 2016 through 2019, the test part of Voxceleb2, FVCCMN, Voices, FBI and LASRS. Details on each of the datasets along with the number of speakers and number of same-speaker (target) and different-speaker (impostor) trials can be found in \ref{sec:appendix_test_data}.

For most of the results we use summary metrics were we average the results over six different groups:
\begin{itemize}[leftmargin=1.5em]\itemsep0em
\item {\bf TRNH}: The six sets composed of speakers held-out from training data. We use these sets for development.
\item {\bf DEV}: The two additional development sets, SRE16 Dev and FVCCMN, corresponding to conditions unseen during training.
\item {\bf VOX}: Voxceleb2 and SITW Eval.
\item {\bf SRE}: SRE16 Eval, SRE18 Eval, and SRE19 Eval.
\item {\bf VOICES}: Voices Eval.
\item {\bf XLAN}: FBI and LASRS.
\end{itemize}

As explained in Section \ref{sec:train_stages}, in order to select a model from a certain training process, after a first \emph{warm up} stage, we start evaluating the models obtained after every mini-batch update. This evaluation is done using the average over the sets included in the TRNH and DEV groups. As we will show in the results, including the two unseen DEV sets when selecting the best model is essential to achieve robustness on the evaluation sets. For each of the eight sets, we also include a version where each segment in each trial is chunked from 4 to 30 seconds (with a uniform distribution in the log domain) to ensure that the selected models are robust on short durations as well. Finally, the performance on the 16 sets (both the full-duration and the short-duration version of the eight sets) is averaged to select the best model. This same averaged metric was used to the select  hyperparameters like the learning rate, architecture, side-information size and all other development decisions. Results on the rest of the datasets were not used during development. When showing results for a certain group we average both over the original sets and corresponding chunked versions between 4 and 16 seconds (again, using a uniform distribution of durations in the log domain). In this case, we use more extreme durations than during development for a more challenging condition as well as additional mismatch with the development scenario. 

Note that chunks (partial speech segments) are used in all stages of training and evaluation. For embedding extractor training we use the standard chunking approach for this purpose, with durations between 2.0 and 3.5 seconds, as explained in Section \ref{sec:appendix_backend_data}. For backend training we either use the full segments or chunks between 4 and 240 seconds. For development we use, for each set, the original full-segment version and a new version created by chunking between 4 and 30 seconds. Finally, for the evaluation of the selected models, we use, for each set, the full-segment version and a new version created by chunking between 4 and 16 seconds. The use of chunks was essential for getting a robust model across durations since most datasets used in our experiments are composed of relatively long segments. 

\subsection{Performance Metrics}
\label{sec:metrics}
In machine learning, there has been much recent interest in measures of the goodness of the calibration of probabilistic classifiers. \citet{guo:17} and many others  that cite them, motivate the need for calibration by the requirement to be able to make good decisions in the face of uncertainty. We agree. Unfortunately, \citet{guo:17} and subsequent publications lose sight of this requirement and they judge calibration with measures that are not designed to reflect decision-making ability. On the other hand, the measurement of calibration via \emph{proper scoring rules} \citep[see][and references therein]{properscoring} is a mostly solved problem. Proper scoring rules quantify the risk (expected cost) of using the outputs of  probabilistic classifiers to make minimum-expected-cost Bayes decisions \citep{brummer_thesis}. As shown in equation~\eqref{eq:BD}, that is exactly how we apply the LLR outputs of our speaker verifier: we use the LLRs to make Bayes decisions. We therefore use a proper scoring rule to evaluate the goodness of our LLRs. The ubiquitous classifier training objective, cross-entropy\footnote{also know as \emph{negative log-likelihood}} is indeed the expected value of a proper scoring rule. In \citep{Brummer:csl06, brummer_thesis} we motivated the use of prior-weighted cross-entropy as a measure of the goodness of binary classifiers that output LLRs. We termed this measure Cllr (cost of LLR). In this work, we show results in terms of this metric. 

Like any other proper scoring rule, this metric is sensitive to both the discrimination and the calibration performance of the system. Specifically, the Cllr is defined as the weighted cross-entropy in equation (\ref{eq:crossent}) with target prior $\pi = 0.5$. We also show results in terms of a weighted cross-entropy with $\pi = 0.01$, which represents applications where the effective prior probability of targets is small. We will call these metrics, Cllr.5 and Cllr.01, respectively. During training we optimize Cllr.01. We also select the best model per run (best seed and epoch) and hyperparameters based on this metric. In our experiments, training and optimizing for this metric resulted in better test Cllr.01, without hurting Cllr.5.

Even a very discriminant system (with low equal-error-rate), can have a high Cllr (for any value of $\pi$) if the calibration is bad. Such a system would lead to bad decisions when thresholded with the theoretical Bayes decision threshold, for some or all of the cost functions that represent the applications of interest. In order to analyze what part of the Cllr is due to miscalibration, we can compute the value for this metric when the test scores are calibrated optimally. This is usually done by using the PAV algorithm \citep{brummer:pav} on the test scores, which finds a non-parametric monotonic transformation that optimizes the value of any proper scoring rule, including Cllr. The optimal value of the Cllr is called minimum Cllr since it is the best Cllr that can be obtained on the test set without changing the discrimination power of the scores. The scores transformed with the PAV algorithm form a useful reference for judging calibration: the relative difference between the actual Cllr and the minimum Cllr indicates what percentage of the Cllr is due to miscalibration of the scores. 

Note, though, that the minimum Cllr obtained with the PAV algorithm can be too optimistic since the transformation that is learned with this algorithm is prone to overfitting when few samples are used to estimate it. An alternative way to obtain a minimum value for the Cllr is to train a linear transform using, for example, linear logistic regression (as explained in Section \ref{sec:global_calibration}) on the test set itself. We have observed in our experiments that the minimum value obtained with such a transformation is almost identical to the one obtained with PAV when the datasets are large, but that the difference is larger when the test sets are smaller (up to 20\% relative difference on test sets with only a few hundred target samples). This suggests that the PAV minimum is too optimistic on those small sets. Hence, in this work we use the best linear transformation obtained with linear logistic regression on the test set to obtain the minimum Cllr. 

Finally, since the Cllr is not as widely used in the speaker verification literature as other metrics, we show summary results using the detection cost function (DCF), which is the average cost on the test data, with a prior probability of same-speaker, $P(H_s)=0.01$, and unity costs, $C(\text{accept}|H_d)=C(\text{reject}|H_s)=1$ (see equation \ref{eq:BD}). That is, we compute  $\text{DCF}= 0.01 P_{miss} + 0.99 P_{fa}$, where $P_{miss}$ is the probability of labeling a same-speaker trial as a different-speaker trial and $P_{fa}$ is the probability of labeling a different-speaker trial as a same-speaker trial. The errors are computed on hard decisions made by thresholding the scores with the threshold that would result in the best expected DCF if the scores were well calibrated (Equation \ref{eq:bayesthr}). 
In contrast with the Cllr, the DCF measures the performance of hard decisions for a single operating point rather than the overall quality of the scores. As for the case of the Cllr, a minimum value of DCF can also be obtained to determine what percentage of the DCF is due to miscalibration. In this case, the minimum is obtained by simply sweeping a threshold and choosing the one that minimizes the DCF.

For a gentle introduction to calibration, the Cllr metric, the PAV algorithm, DCF, and optimal decisions, please see \citep{van2007introduction}.

\subsection{Training and Architecture Hyperparameters}
\label{sec:hyperparameters}

Initialization parameters for the LDA and PLDA stages are estimated using all the training data when using full segments and using a single chunk per segment when using chunks.
To initialize the calibration stage we do not use all the available training data since that would be computationally costly and also unnecessary given the small number of parameters to optimize. Hence, we simply select 1000 speakers and create trials, using one of the two procedures described in Section \ref{sec:batches}.

We train the models using the procedure described in Section \ref{sec:train_stages} with a fixed learning rate of 0.0005 during the first stage which consists of 12000 mini-batches. Then, for the second stage when we start testing performance after every update, we increase the learning rate to 0.001, which improved performance over using the smaller learning rate for this stage. We do this over 3000 mini-batches. Finally, in the third stage we fine tune the best model from the second stage with a smaller learning rate of 0.00001 over 100 mini-batches. We use a batch size of 2048, which resulted in better performance than the smaller sizes we used in our previous papers. We use L2 regularization on all parameters, with weights tuned on the development data, and clip the gradients to a maximum norm of 4.0. 

Unless otherwise noted, we use an LDA dimension of 300, a dimension for $m$ (equation \ref{eq:m}) of 200, dimension of 6 and identity transform as $f$ for $z$ (equation \ref{eq:z}), and duration features given by equation (\ref{eq:elb}) with $c=30$ and $s=2$. These default parameters for the architecture and those for the training procedure described above are provided in configuration files in the example directory in \url{https://github.com/luferrer/DCA-PLDA}.

\section{Results}
\label{sec:results}

In this section we include various results and analysis comparing the standard PLDA backend with different configurations of the proposed backend for different training methods. 
We use the following naming convention for the systems. Systems for which the LDA and PLDA parameters are frozen at their standard values are called PLDA. Systems where the LDA and PLDA stages are trained discriminatively after initialization with the standard values are called D-PLDA. For both cases, the calibration stages are indicated as suffixes: DD for duration-dependent calibration, SD for side-information-dependent calibration, and DSD for duration- and side-information-dependent calibration. The full system proposed in this paper is then called D-PLDA-DSD. We also call this system DCA-PLDA, which is easier to pronounce and remember. We use the harder-to-pronounce D-PLDA-DSD name when convenient for indicating the difference between systems under comparison.

\subsection{Training Setup}
\label{sec:res_train}
\begin{figure}
\centering
\includegraphics[width=0.7\linewidth]{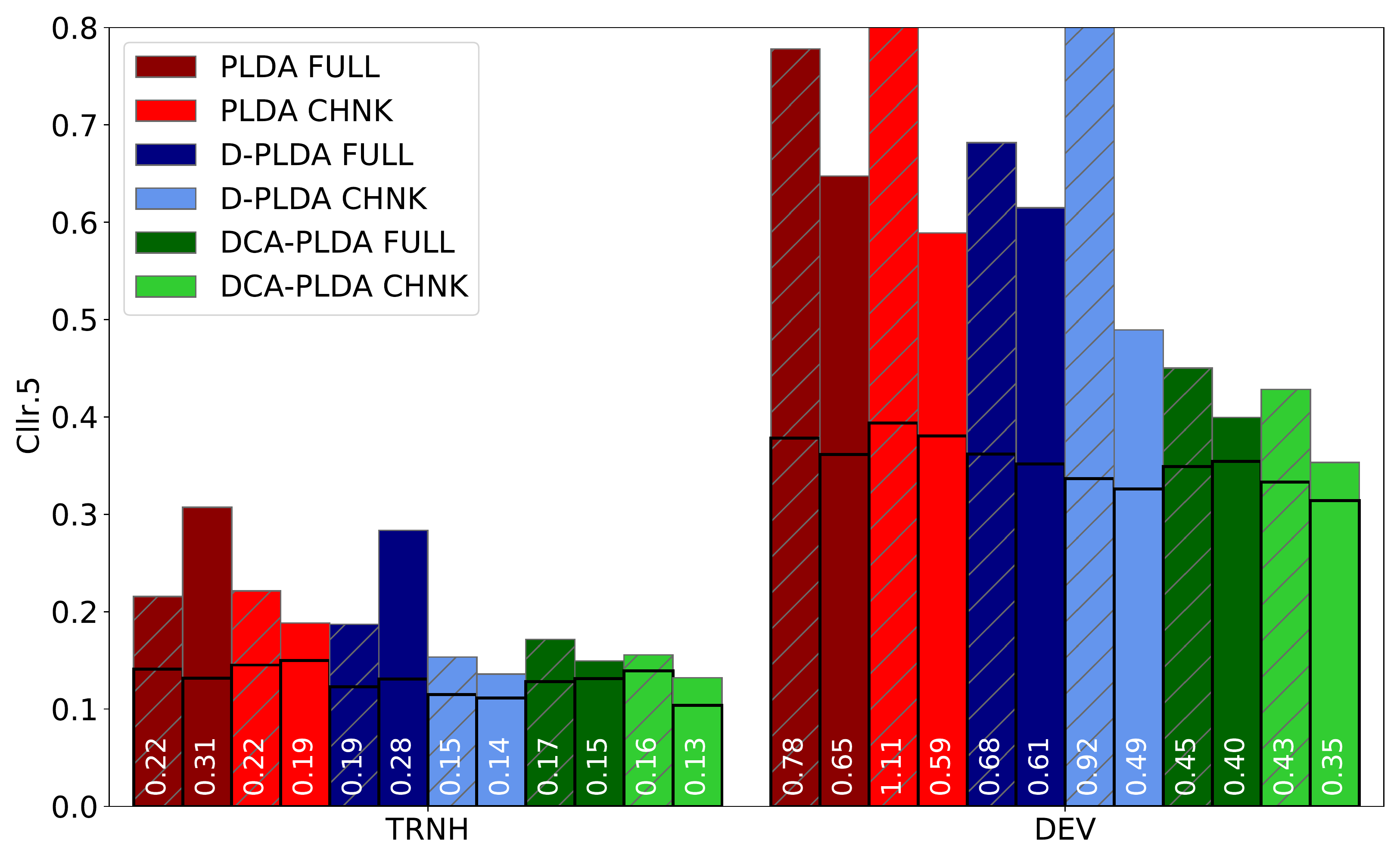}
\caption{Average actual and minimum (black lines) Cllr with $\pi=0.5$ for the TRNH and DEV sets for standard PLDA, D-PLDA, and DCA-PLDA backends, using full waveforms (FULL) or chunks (CHNK) for training, for two speaker weighting approaches, flat (striped bars) and balanced-by-domain (solid bars). The y-axis is cut at 0.8 to focus on the differences between the best performing systems. The white numbers within the bars indicate the height of the bars.}
\label{fig:init}
\end{figure}

In this section we show support for two important decisions made in terms of training setup: the initialization procedure and the use of training data as full or chunked waveforms. 
Figure \ref{fig:init} shows results for the standard PLDA backend, a D-PLDA backend obtained by replacing the duration and side-information dependent calibration stages in DCA-PLDA with a global calibration stage, and the full proposed backend, DCA-PLDA, with both calibration stages. Results are shown on the two groups of sets that are used for development, TRNH and DEV. 

We compare results using two different approaches for initialization and batch generation. In one case, which we call \emph{flat}, all training samples are considered equally important. The PLDA model is learned setting all speaker weights in equations (\ref{eq:mu0}), (\ref{eq:B0}), and (\ref{eq:W0}), and during EM iterations equal to 1.0. Further, when generating batches for D-PLDA or DCA-PLDA,  training speakers are selected randomly with equal probability. Hence, in this case, the most frequent domain, which is VOX, containing half of the training speakers, dominates the training process. In the other case, which we call \emph{ balanced-by-domain}, we set the speaker weights for PLDA initialization and EM iterations to be the inverse of the number of speakers for their domain (given the way we define the domains, each speaker only appears in a single domain). Further, when generating batches for D-PLDA or DCA-PLDA we use the second approach described in Section \ref{sec:batches}, where each batch contains the same number of speakers for each domain. Hence, in this case, all domains are equally represented during training.

We show results when training with full waveforms and using chunks (see Section \ref{sec:backend}). Arguably, the DCA-PLDA with duration-dependent calibration model should only be trained with chunked data, since otherwise it would not be able to separate the effect of duration and side-information, given that the duration distribution for the original waveforms is highly domain dependent. On the other hand, after chunking, all domains have a uniform log-duration distribution between 4 and 240 seconds, which should better allow the model to separate the effect of duration and side-information on the calibration parameters.

We can see that for all three systems, the best training approach is to use chunks and balance the data by domain. For the rest of the experiments in this paper, chunks and balanced-by-domain weights and batches are used to train all systems.

Finally, we also tried random initialization for DCA-PLDA, using a normal distribution centered at 0.0 with standard deviation of 0.5. These results are not included to reduce clutter in the figure; they are worse than the ones shown here, especially on the DEV sets where random initialization led to up to 60\% worse results for the balanced-by-domain case. This highlights the importance of the initialization process described in Section \ref{sec:init} for getting optimal performance with the proposed approach.

\subsection{Learning Curves}
\label{sec:learning_curves}

\begin{figure}
\centering
\includegraphics[width=0.34\linewidth]{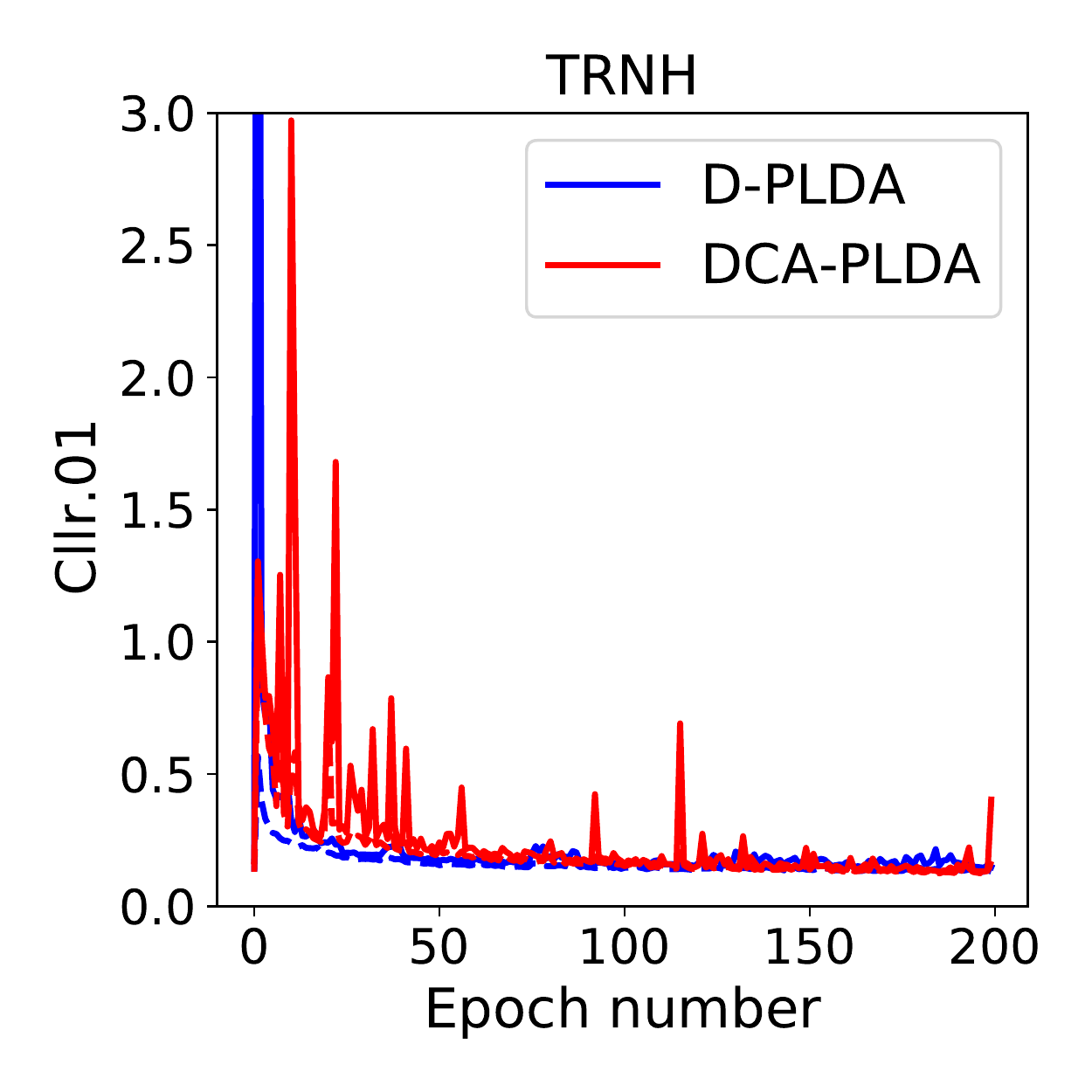}
\includegraphics[width=0.34\linewidth]{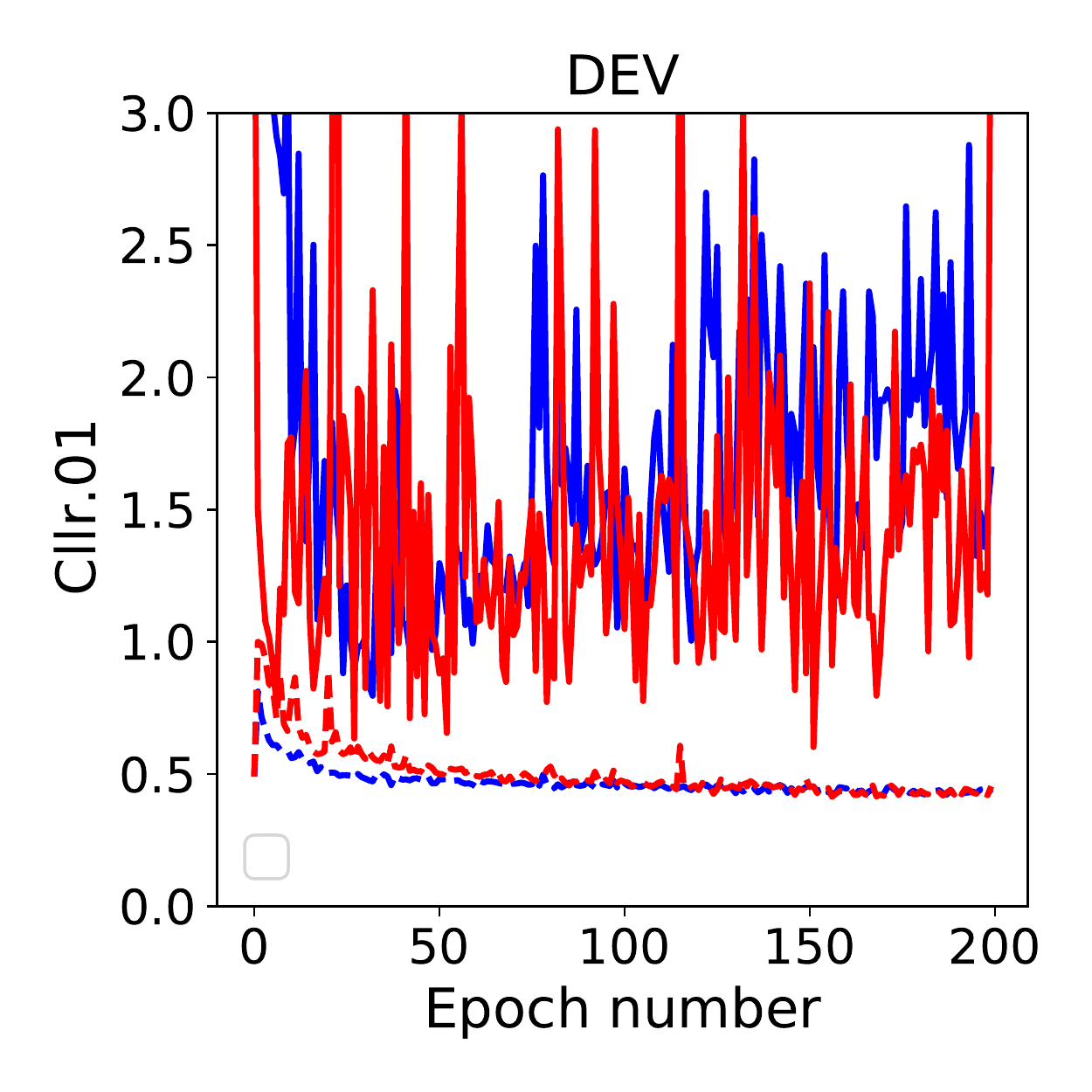}
\includegraphics[width=0.28\linewidth]{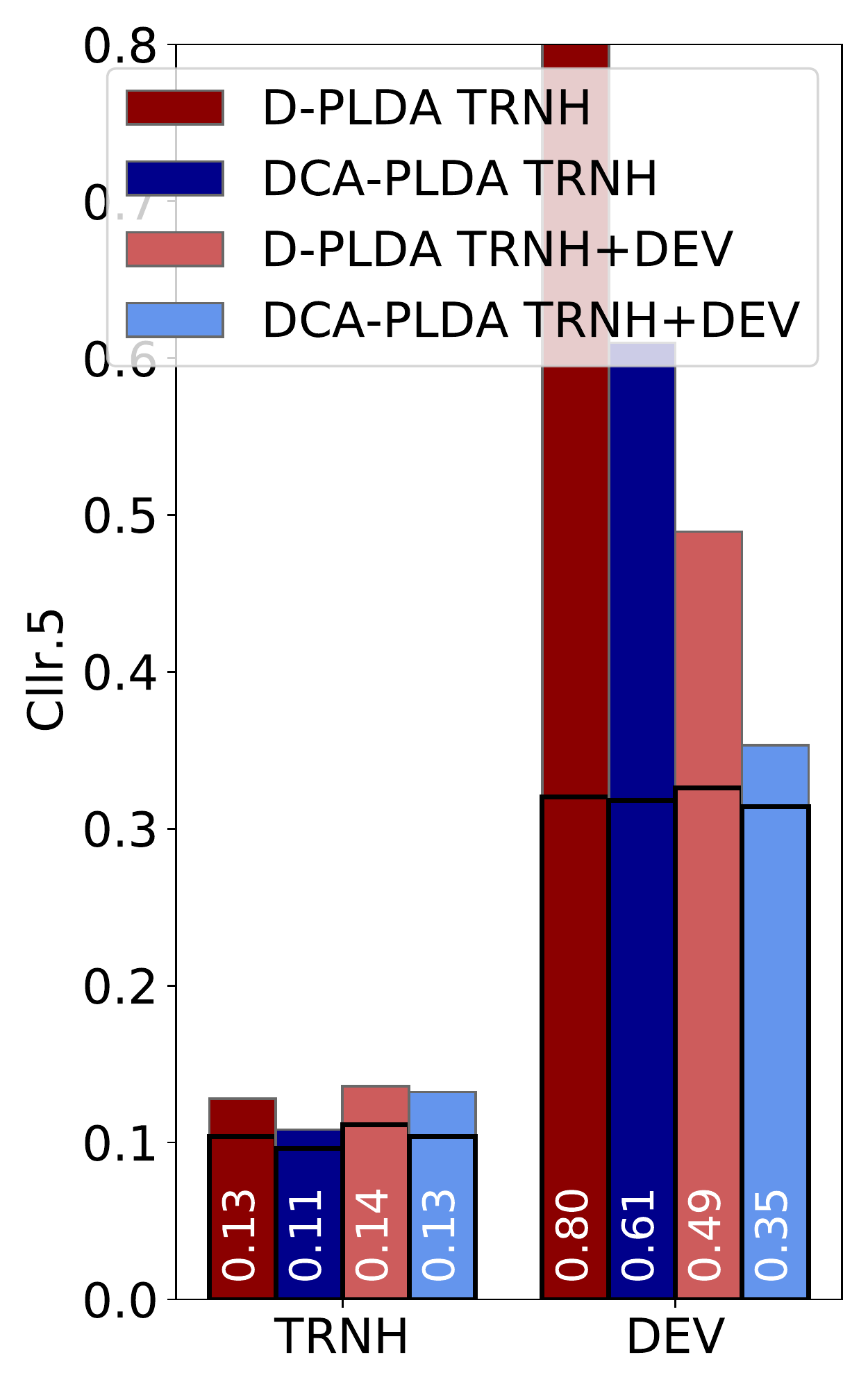}
\caption{{\bf Left and center:} Cllr.01 (the training objective function) for the first 200 epochs composed of 10 mini-batches each on the TRNH and DEV groups for two systems, D-PLDA and DCA-DPLDA. Solid curves correspond to actual Cllr.01, while dashed curves correspond to the minimum Cllr.01. {\bf Right:} Cllr.5 on TRNH and DEV for the D-PLDA and DCA-PLDA models selected using only TRNH sets, or using the average over all TRNH and DEV sets.}
\label{fig:learncurves}
\end{figure}

The left and center plots of Figure \ref{fig:learncurves} shows the loss used for training (Cllr.01) for the TRNH and DEV groups, for two systems D-PLDA and DCA-PLDA. We show the actual (solid) and minimum (dashed) Cllr.01 for every epoch composed of 10 mini-batches each. The loss at iteration 0 corresponds to the model initialized as described in Section \ref{sec:init}. We can see that for the TRNH sets, which correspond to unseen speakers from conditions seen on the training data, after the first few mini-batches, the learning curves are smooth and stabilize at a low value, specially for the simpler D-PLDA model. On the other hand, on the DEV sets, which correspond to unseen conditions, the actual Cllr is quite unstable. Interestingly, this does not happen for the minimum Cllr, which changes relatively little across epochs. The minimum Cllr in these curves is obtained as the average Cllr over all test sets in each group, after transforming the scores with an affine function obtained with linear logistic regression trained on each test set. The fact that these curves are smooth implies that the changes in parameter values that occur between epochs result in similar shift and scaling of the scores for all scores from each test set, so that the discrimination performance within each set (and, hence, minimum Cllr) is not affected by these changes.
On the other hand, the shift and scaling of scores that occurs across epochs greatly affects calibration. This observation highlights the importance of using actual instead of minimum loss as a metric during development.

The right plot in Figure \ref{fig:learncurves} show the results obtained on TRNH and DEV sets for the D-PLDA and DCA-PLDA models selected using only TRNH sets or using the average over all TRNH and DEV sets (our default approach). As we can see, the performance on TRNH sets is better  when using only TRNH sets for selection, but the performance on DEV sets is extremely bad for those models compared to the ones selected using the average over both TRNH and DEV sets. We pay a very large price on unseen conditions when selecting the optimal model on seen conditions. 
For this reason, we use the strategy described in Section \ref{sec:train_stages} to select the best model for a certain run based on the average performance over the TRNH and DEV sets. As we will see, the model selected this way shows good generalization on unseen datasets.  

We have tried several approaches to prevent the DEV performance from changing so drastically between one mini-batch and the next, including using slower learning rates or different learning rate schedules, higher L2 regularization coefficients, and regularizing the output of the system as proposed by \citet{pezeshki2020gradient}. While some of these approaches succeeded in taming the learning curve for the DEV sets, they all resulted in worse final performance.

Finally, we tried adding the two DEV sets to the training data as two separate training domains, one for each set. This led to a degradation in performance on the evaluation sets for both systems. It appears to be important for those two sets which, as shown in this section, are essential for choosing a robust model, to be held-out from training.

\subsection{Duration-Dependent Calibration Stage}
\begin{figure}
\centering
\includegraphics[width=0.27\linewidth]{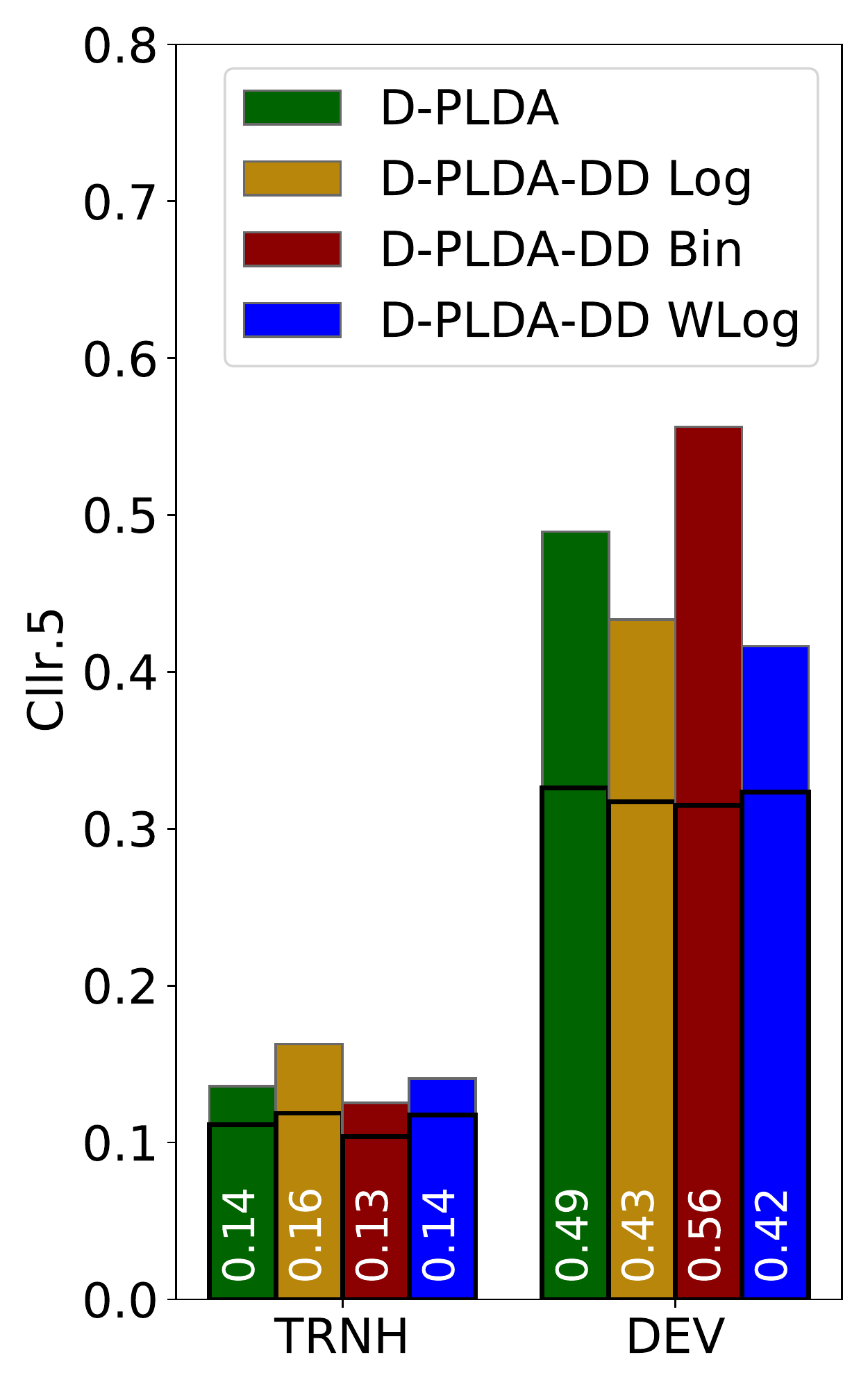}
\includegraphics[width=0.72\linewidth]{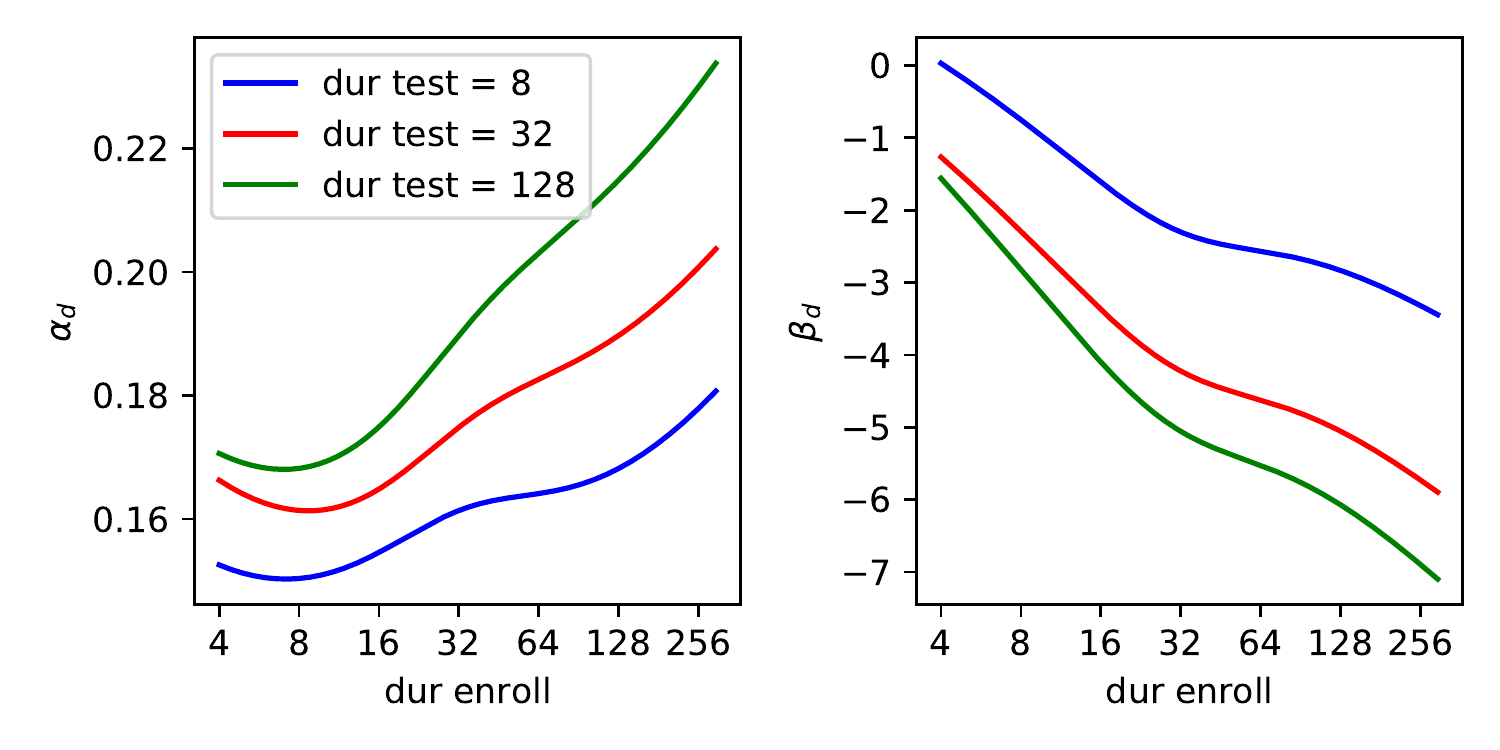}
\caption{{\bf Left:} Average actual and minimum (black lines) Cllr with $\pi=0.5$ for TRNH and DEV sets for a plain discriminative PLDA (D-PLDA) system with global calibration and three duration-dependent-only DCA-PLDA backends (D-PLDA-DD), using different duration features. {\bf Right:} Values of $\alpha_d$ and $\beta_d$ obtained by the D-PLDA-DD WLog model as a function of the durations of the two sides of a trial (dur enroll and dur test).}
\label{fig:dur}
\end{figure}

As described in Section \ref{sec:formulation}, we explored different ways to encode duration information into features that can then be used to obtain condition-dependent calibration parameters using the polynomial forms in Equations (\ref{eq:alpha_d}) and (\ref{eq:beta_d}). These features are given by the logarithm (Log, equation \ref{eq:el}), by binning and one-hot encoding (Bin, equation \ref{eq:eb}), and by a windowed version of the logarithm (WLog, equation \ref{eq:elb}). The left plot in Figure \ref{fig:dur} compares the results for a simple D-PLDA with global calibration, and three D-PLDA systems using only the duration-dependent calibration stage, each with a different set of duration features. For the Bin features we use thresholds of 8, 16, 32, 64 and 128, which are equally spread in the logarithmic domain. For the WLog features, we use $c = 30$ seconds and $s=2$. These values were lightly tuned on TRNH+DEV sets. The side-information-dependent calibration stage is disabled for the experiments in this section to study the effect of duration-dependent calibration alone.  

We can see that on the TRNH sets  only the binning approach gives a modest gain over not using a duration-dependent calibration stage. On the other hand, for the DEV sets, this approach for creating duration features results in a degradation in performance. It appears that the flexibility of this model allows it to overfit the training conditions. Overall, the WLog approach gives the best trade-off across sets. This is the setup we use for the rest of the experiments in this paper. As we will see in our final results, the gain observed on DEV sets with respect to D-PLDA  when using this duration-dependent approach generalizes to the evaluations sets, resulting in larger gains on some conditions.

The center and right plots in Figure \ref{fig:dur} show the values of $\alpha_d$ and $\beta_d$ obtained with the WLog model as a function of the original duration for the enrollment and test sides of a trial. We can see that, while there is some displeasing non-monotonicity in $\alpha_d$ at the lower end of the durations, the trends are reasonable: larger durations correspond to larger values of $\alpha_d$ which implies larger values of the resulting LLR (i.e., increased confidence). Also, $\alpha_d$ is always positive, meaning that the sign of the scores that come out of the PLDA stage are never reversed. 

\subsection{Side-Information-Dependent Calibration Stage}

We explored three different forms for the $f$ function in equation ($\ref{eq:z}$), identity, softmax and log-softmax (see Section \ref{sec:formulation}). The identity gave a slightly better result than the other two, though all three were similar in performance. We also explored different dimensions for the $z$ vector, selecting the value that gave the best average performance on TRNH plus DEV sets, which was a dimension of 6. Nevertheless, any values between 4 and 8 gave similar performance. This optimization was done on the full model with both calibration stages, using the duration-dependent calibration stage with the parameters selected as described in the previous section. For the rest of the experiments in this paper we use the identity function as transform and 6-dimensional $z$ vectors.

\begin{figure}
\centering
\includegraphics[width=1.0\linewidth]{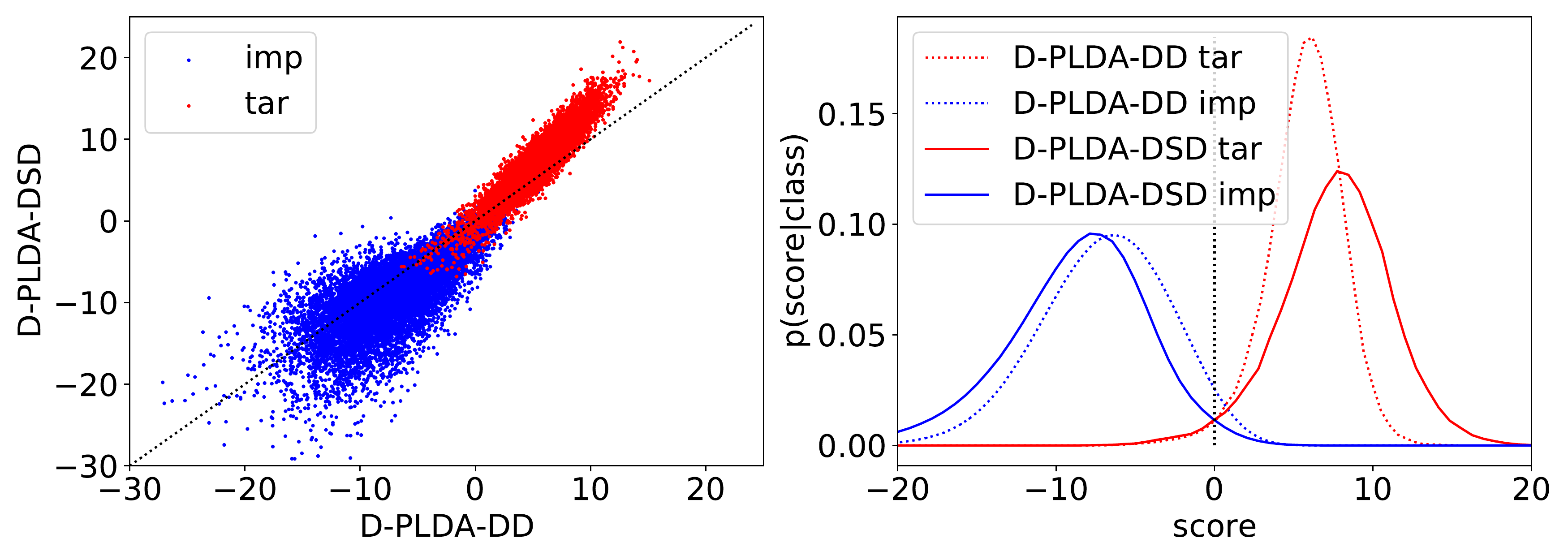}
\caption{Scatter plot and distributions of the output scores ($l$ in Figure \ref{fig:architecture}) from the D-PLDA-DD and D-PLDA-DSD systems on the Vox2-Tst set.}
\label{fig:sideinfo}
\end{figure}

Analyzing the values of $\alpha$ and $\beta$ for the side-information-dependent calibration stage as we did for the duration-dependent stage is not possible since the side-information vector is not unidimensional. 
One way to visualize the effect of the side-information dependent calibration as a whole is to compare the scores from the D-PLDA-DD selected in the previous section, with the scores from the full system, D-PLDA-DSD. Figure \ref{fig:sideinfo} shows the scatter plot of the two scores and their distributions for the Vox2-Tst dataset. In this case, the D-PLDA-DD system has a Cllr.5 of 0.11, while the D-PLDA-DSD system has a Cllr of 0.08 (see Figure \ref{fig:evaldata}). This improvement is seen in the fact that the target and impostor distributions for the D-PLDA-DSD system cross at zero, which is a property of well-calibrated LLRs \citep{van2013distribution}, while the ones for the D-PLDA-DD system do not. 

In the scatter plot we can see that, the correlation between target scores for both systems is strong, while for the impostor scores it is much weaker.
This suggests that the side-information vectors $z$ are not only reflecting speaker-independent information (the $h$ variables from Section \ref{sec:condition}) since that information should be similar for target and impostor trials within this set and would result in strongly correlated scores for both classes, but also some rough speaker-dependent information (the $g$ variables from Section \ref{sec:condition}) which would be the same for both sides of a target trial but differ in impostor trials. We can also see that the D-PLDA-DSD system is more confident: the average slope in the scatter plot is greater than one and the target and impostor means move further apart.    

\subsection{Effect of Joint Training}

In this section we show the effect of discriminatively and jointly training all parameters in the model versus training only the calibration parameters discriminatively using generatively-trained LDA and PLDA models for score generation. For each case, training the full model (D-PLDA) or training only the calibration stages (PLDA), we run four options: global calibration (no suffix), side-information-dependent calibration (SD), duration-dependent calibration (DD), and side-information- and duration-dependent calibration~(DSD).  

\begin{figure}
\centering
\includegraphics[width=1.0\linewidth]{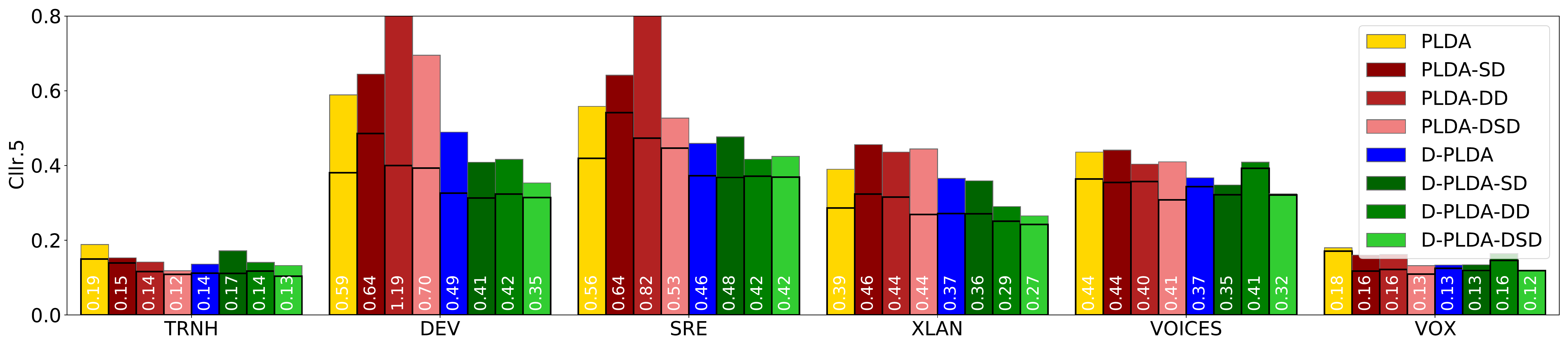}
\caption{Comparison of PLDA and D-PLDA systems with and without duration- and side-info-dependent calibration stages. Note that the D-PLDA-DSD system is the full DCA-PLDA system proposed in this paper.}
\label{fig:architectures}
\end{figure}

Figure \ref{fig:architectures} shows the results for those eight systems on all six groups described in Section \ref{sec:evaldata}. Results show that joint training of all the model's parameters gives a substantial and consistent gain across all architectures. That is, the D-PLDA version is always better than its corresponding PLDA version, in some cases by a very large margin. Notably, condition-dependent calibration only gives consistent gains over global calibration if the full model is trained jointly. 

The one exception to these observations is on the TRNH sets, where the PLDA versions work quite well. It is important to note that this is the most common scenario in speaker verification papers, where the PLDA model is trained on data that is matched to the evaluation data. In this scenario, the jointly trained models do not offer a consistent advantage; D-PLDA is better than PLDA only when global calibration is used. This highlights the importance of evaluating performance on mismatched conditions. While all models are somewhat similar on conditions matched to the train conditions, they are substantially different on unseen conditions. In particular, the D-PLDA-DSD model gives the best trade-off across sets. 

\subsection{Variation with Seed}

As explained in Section \ref{sec:train_stages}, all results shown in this paper correspond to the best model in terms of average Cllr.01 over the TRNH and DEV sets, obtained out of 20 runs with different random seeds. For our proposed system, D-PLDA-DSD, the median performance across the 20 seeds (approximated by the average Cllr.01 performance of the seed number 10 after sorting them based on performance) and the worst performance are 6\% and 14\% worse than the best performance, respectively, while for D-PLDA these numbers are 22\% and 38\%. Hence, despite the fact that our model is slightly more complex than D-PLDA in terms of number of parameters, it gives a more stable performance across seeds. A similar trend is observed for the results in Figure \ref{fig:dur} where the best model, D-PLDA-DD WLog, has smaller relative range of performances across seeds than the other systems in that figure. These results suggest that better models have an easier time reaching a good set of parameters, regardless of the seed.

\subsection{Comparison with a System Trained on VOX-only Data}
\begin{figure}
\centering
\includegraphics[width=0.7\linewidth]{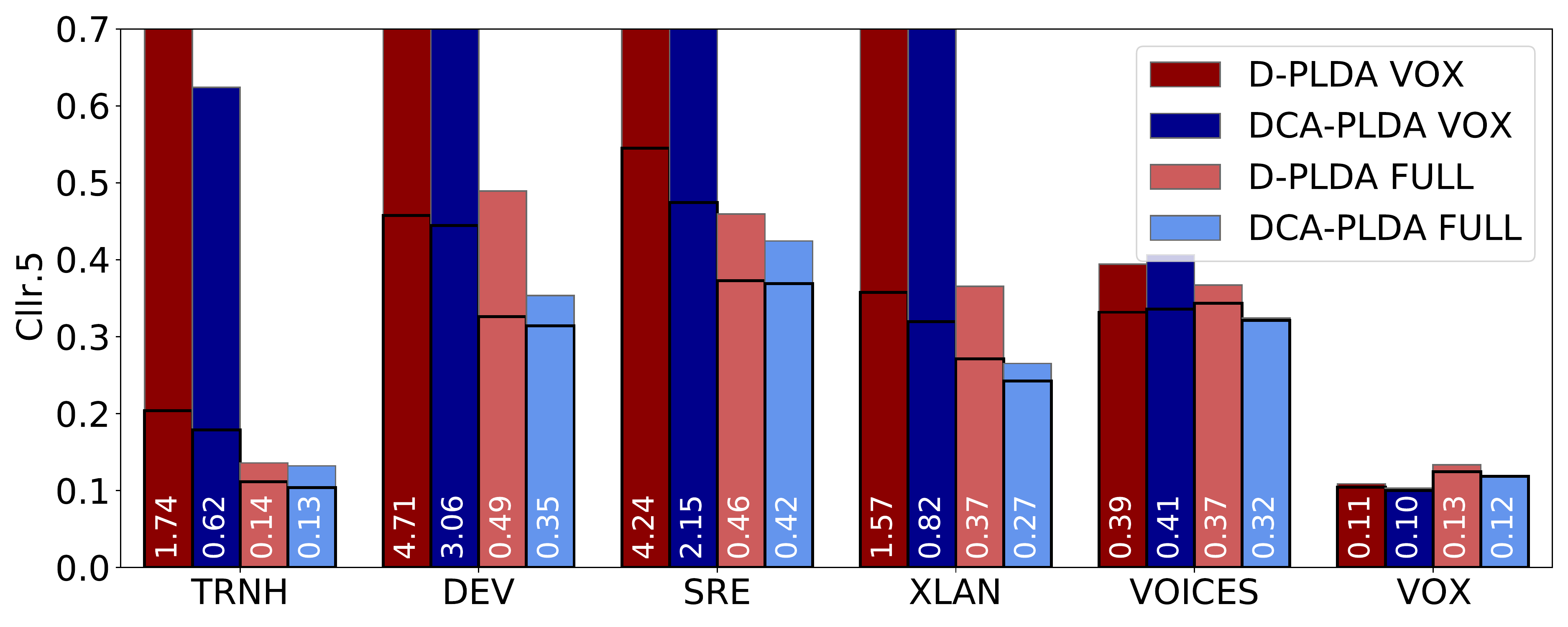}
\caption{Comparison of D-PLDA and DCA-PLDA (a.k.a, D-PLDA-DSD) systems (1) training the models with all training domains and selecting the best epoch and seed using TRNH and DEV sets (FULL), and (2) training the models only with the VOX domain and selecting the best epoch and seed using the held-out VOX data, one of the six sets in the TRNH group (VOX).}
\label{fig:res_vox}
\end{figure}

In this section, we select two systems D-PLDA and D-PLDA-DSD and compare their performance on all evaluation groups when the systems are trained in two ways: (1) with the default approach, using all training domains for training and selecting the best epoch and seed using TRNH+DEV sets, and (2) using only the VOX training domain and selecting the best epoch and seed using only the heldout part of the VOX training data. The latter would be the standard approach when developing a system exclusively for VOX-like data. Figure \ref{fig:res_vox} shows that training the model using a variety of conditions is 16\% worse on VOX sets (SITW-Eval and Vox2-Tst) compared to using only VOX data. Yet, this loss is small compared to the gain obtained on all other sets. Again, as we saw in Section \ref{sec:learning_curves} with the issue of model selection, a large price is paid on unseen conditions in exchange for a moderate gain on seen conditions. This again highlights the importance of evaluating systems on a variety of conditions when trying to develop a system that is robust to unseen conditions.

\subsection{Comparison with Adaptive S-Norm}
\begin{figure}
\centering
\includegraphics[width=0.7\linewidth]{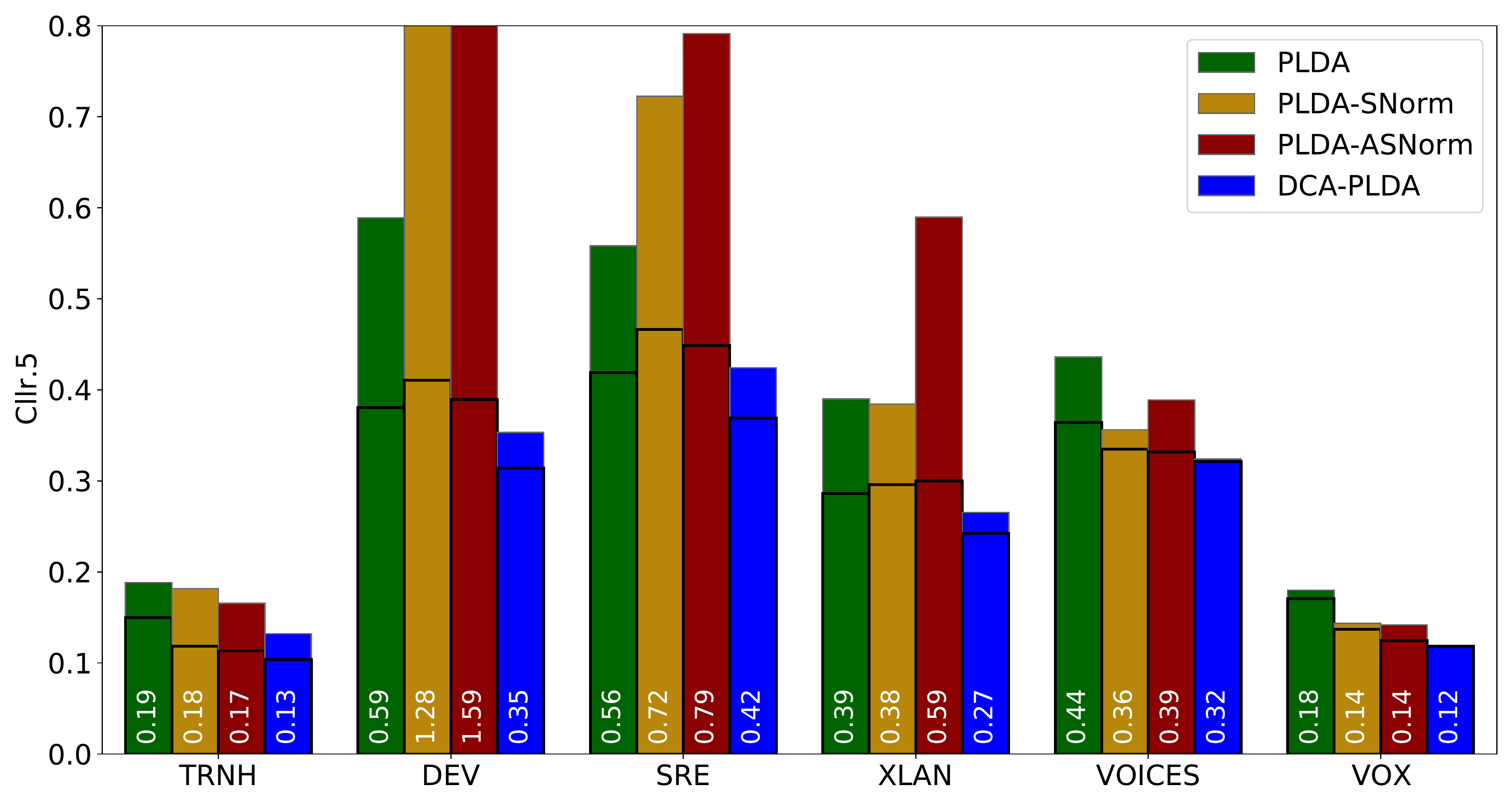}
\caption{Comparison of PLDA, PLDA with S-Norm and AS-Norm, and DCA-PLDA. Training is done using chunks and the balanced-by-domain approach.}
\label{fig:snorm}
\end{figure}

As mentioned in Section \ref{sec:prior_work}, adaptive normalization could potentially achieve our goal of mitigating the effect of the samples' conditions in the scores. In this section we show results comparing one such approach to our proposed method. We implement the adaptive symmetric normalization (AS-Norm) method described in \citep{karam2011towards}, called AS-Norm1 in \citep{matejka2017analysis}. In this approach, the score for each trial is computed as the average of a test-normalized and an enrollment-normalized score. The test-normalized score is computed as $(s-m_t)/s_t$, where $s$ is the trial's score, and $m_t$ and $s_t$ are the mean and standard deviation of the top $N$ scores obtained on trials composed of the test sample versus a set of cohort samples. The enrollment-normalized score is computed similarly, using the enrollment sample. In our experiments, the cohort samples are extracted from the training data selecting 1000 random samples from each domain. Calibration is done as for the unnormalized case, using trials created from 1000 of the training speakers, as described in Section \ref{sec:hyperparameters}. We tuned $N$ on the TRNH group to a value of 900. The optimal value over the DEV and TRNH groups together is 6000, which corresponds to plain S-Norm. Figure \ref{fig:snorm} shows the comparison between PLDA without normalization, PLDA with S-Norm and with AS-Norm, and the proposed DCA-PLDA method.

The figure shows that S-Norm and AS-Norm work well on data that is well-matched to the training data --- the TRHN and VOX groups. In these cases, the minimum Cllr and actual Cllr improve over that of PLDA without normalization. Nevertheless, DCA-PLDA still outperforms PLDA with normalization in these conditions. Notably, both normalization approaches fail on unseen conditions in terms of actual Cllr. Hence, we again see that the performance on matched conditions is a poor predictor of the performance on unseen conditions.

\subsection{Final Results on Evaluation Data}

In this section we present complete results on all sets described in Section \ref{sec:evaldata}, excluding the ones used for development. We show results in terms of Cllr.5, which we have been using to show most results in this section, Cllr.01, which we use during system training and model selection, and DCF, as described in Section \ref{sec:metrics}. We compare four systems, two systems that we consider our baseline, PLDA and D-PLDA, the proposed D-PLDA system with duration-dependent calibration and the full D-PLDA-DSD system with both calibration stages (which we have also called DCA-PLDA). 

\begin{figure}
\centering
\includegraphics[width=0.9\linewidth]{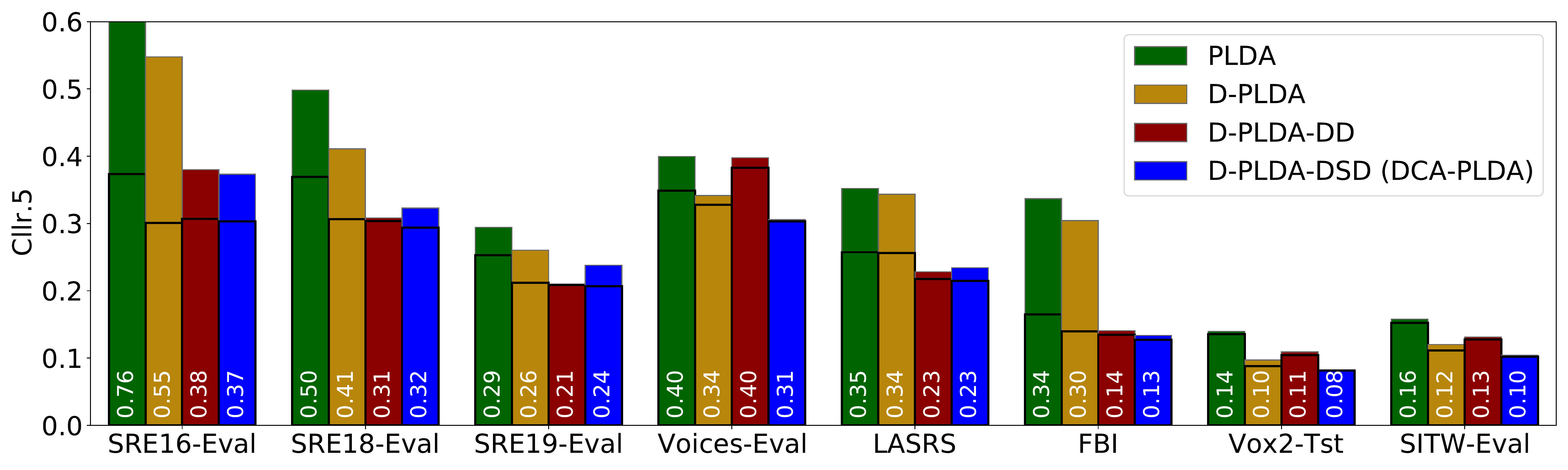}
\includegraphics[width=0.9\linewidth]{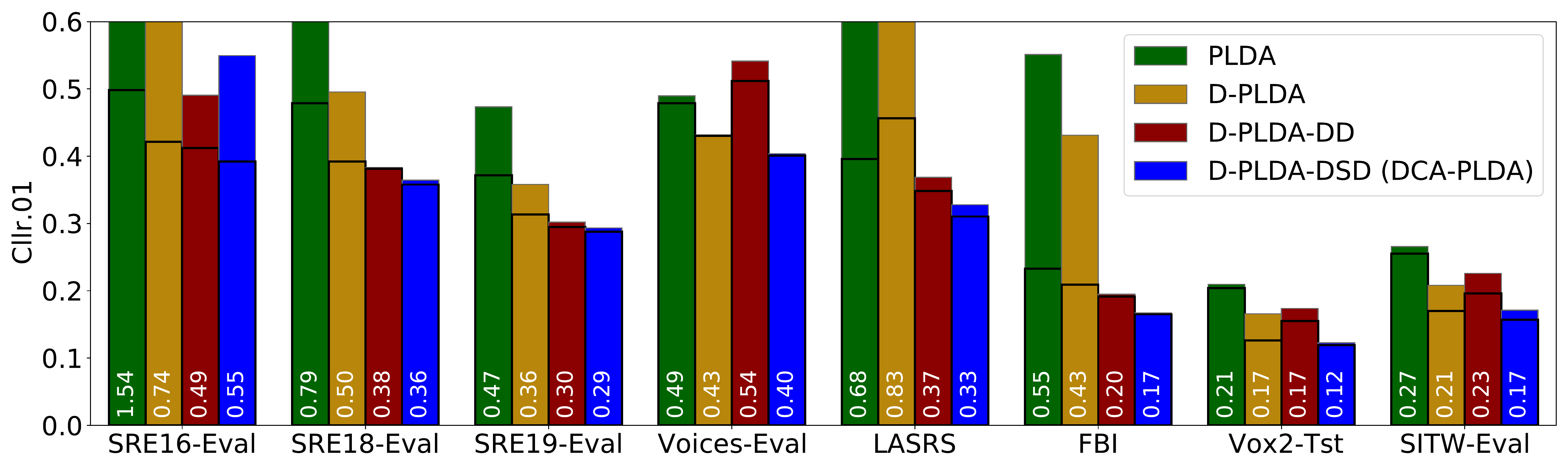}
\includegraphics[width=0.9\linewidth]{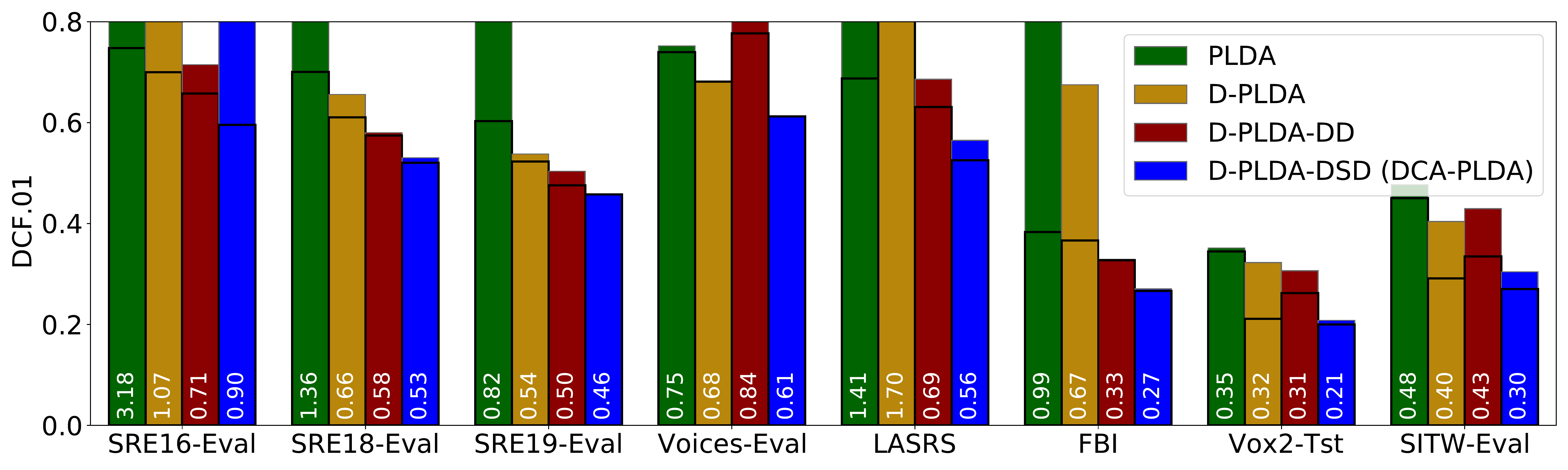}
\caption{Comparison of PLDA, D-PLDA baseline systems with two proposed systems, D-PLDA-DD and D-PLDA-DSD (also called DCA-PLDA) on all evaluation sets for the two Cllr metrics and DCF with an effective probability of target of 0.01.}
\label{fig:evaldata}
\end{figure}

Figure \ref{fig:evaldata} shows that the D-PLDA-DSD system outperforms the two baseline systems, PLDA and D-PLDA, on all datasets without exception. Note that these baseline systems are better than in a standard implementation since we train the models using chunked waveforms and using weights to balance data by domain, two strategies which, as we saw in Section \ref{sec:res_train} (Figure \ref{fig:init}), give an advantage over the standard approach of training with full waveforms and equal weights for all speakers. Comparing the D-PLDA-DD and the full D-PLDA-DSD systems we see that, in most cases the latter is better with some exceptions on SRE data. Overall, D-PLDA-DSD gives the best trade-off across datasets.

\begin{figure}
\centering
\includegraphics[width=1.0\linewidth]{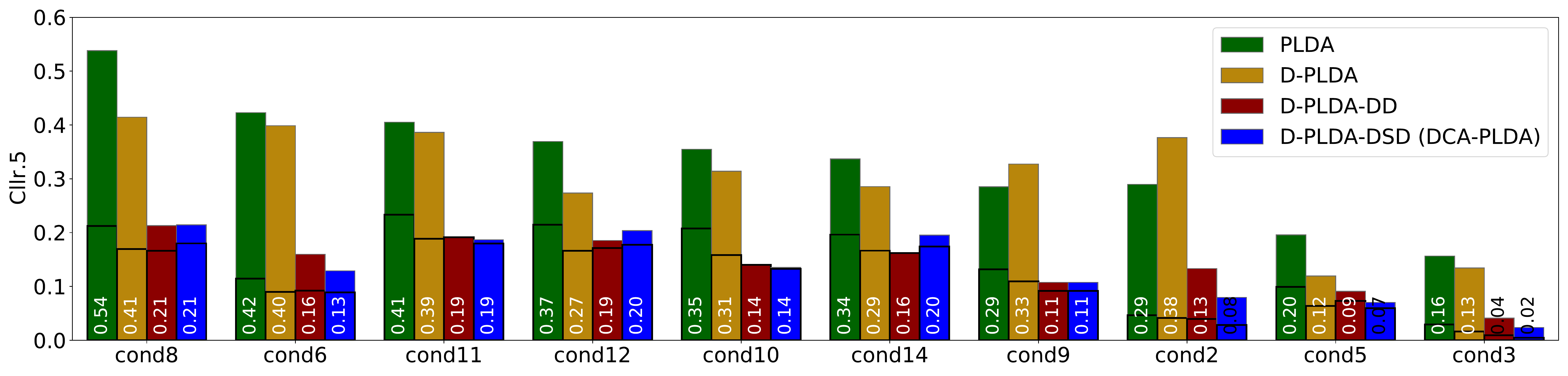}
\includegraphics[width=1.0\linewidth]{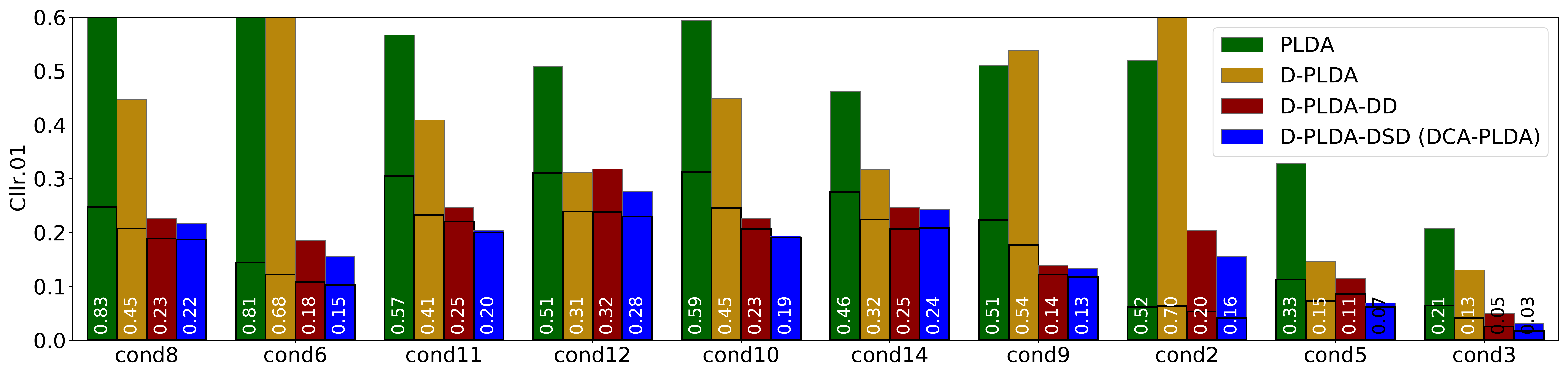}
\includegraphics[width=1.0\linewidth]{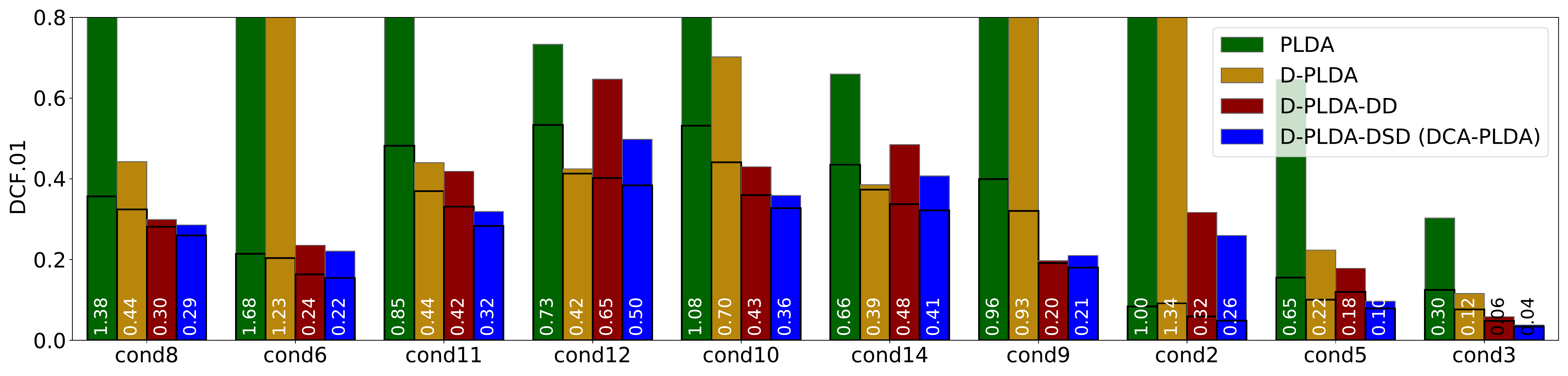}
\caption{Same as Figure \ref{fig:evaldata} on FBI subsets.}
\label{fig:fbidata}
\end{figure}

Figure \ref{fig:fbidata} shows the results for the same four systems in Figure \ref{fig:evaldata} on subsets of the FBI data, which was specifically designed by the FBI for work on calibration. We see that the DCA-PLDA system is remarkably better than both PLDA and D-PLDA on this dataset. In most cases the largest gain is due to the duration-dependency of the calibration stage. Yet, in the easier conditions (cond2, cond5, cond3) the additional side-information-dependent calibration stage gives a further gain. 

\subsection{Model Size and Run Time}

With the configuration used for our experiments, the number of parameters for the PLDA and D-PLDA systems is $334\,203$. For D-PLDA-DD, the number of parameters is $334\,223$ --- only 20 parameters more than PLDA, corresponding to the $\Lambda$, $\Gamma$ and $c$ parameters in Equations (\ref{eq:alpha_d}) and (\ref{eq:beta_d}). Finally, for D-PLDA-DSD, the number of parameter is $438\,187$ --- an increase of $103\,964$ over D-PLDA-DD, corresponding to the parameters in Equations (\ref{eq:alpha_s}) through (\ref{eq:m}). Hence, there is 31\% increase in model size for D-PLDA-DSD with respect to PLDA.

In terms of run time, scoring the full matrix of trials for Vox2-Tst (i.e., all 24 million trials that can be formed by pairing the 4903 samples in the set) takes 1.2 seconds for PLDA and D-PLDA systems (these two approaches are identical in terms of operations needed for evaluation), 1.9 seconds for D-PLDA-DD and 2.9 seconds for D-PLDA-DSD, on an Intel Xeon 2.4GHz CPU. Hence, the run time for the proposed approach is 2.4 times that of the standard PLDA approach. On a smaller set like SITW-Eval, composed of 1.4 million trials, the relative difference is similar, with PLDA and D-PLDA taking about 0.08 seconds and D-PLDA-DSD taking 0.18 seconds.

\section{Conclusions}

We presented a novel backend approach for speaker verification which consists of a series of operations that mimic the standard PLDA-backend followed by calibration. The parameters of the model are learned jointly to optimize the overall speaker verification performance of the system, directly targeting the loss function of interest in the speaker verification task. In order to achieve good generalization in terms of discrimination and calibration across varying conditions, we introduced a duration-dependent calibration stage followed by a side-information dependent calibration stage, where the side-information vector is meant to represent the hidden condition factors. The side-information vector is derived from the input embeddings and is learned jointly and discriminatively with the rest of the model, without the need for external condition labels. 

The main goal of this work was to develop a speaker verification system that can be used out of the box on a large variety of conditions without need for adaptation data and without explicit knowledge of the test conditions. To this end, the model was trained on a large dataset including a variety of acoustic conditions and languages. We compared our proposed approach with three baselines trained on the same large dataset: a standard PLDA backend, a PLDA backend with adaptive score normalization, and a discriminatively trained PLDA-like backend, all with global calibration. We found that the proposed backend with condition-aware calibration stages gives remarkable improvements over the baseline systems on a wide variety of test conditions, reaching up to 85\% relative improvement in terms of actual Cllr with respect to the standard PLDA backend. Despite these large gains, a relatively large difference between the minimum and the actual Cllr is still observed for some conditions indicating that calibration is not always perfect. This, of course, is expected since the training data for the model cannot possibly cover all conditions that can be encountered during testing. Yet, even in those cases, the proposed system outperforms the baseline approaches, with Cllr values that are always better than those of an uninformative system, something that cannot be said of the PLDA system with global calibration.

The code for training and evaluating the proposed backend was released as an open-source repository, along with a recipe using pre-computed embeddings, hopefully making it simple for researchers and developers to try the approach. 
The main information needed to train the model is, as for the standard PLDA backend, the speaker identity label for each of the training samples.
Unlike for PLDA, though, DCA-PLDA also requires session identifiers, which are used to avoid creating same-session trials for model training, which could result in suboptimal performance. Fortunately, this information is provided in most publicly available speaker recognition datasets.

The biggest challenge when training the proposed model is the selection of appropriate development data. In our results, the performance of the model across training epochs is quite variable when testing on conditions that are not perfectly matched to the training data. Hence, when the goal is to obtain a model that performs well on unseen conditions, it is essential to select the best model (epoch and seed) using held-out data from conditions that are not perfectly matched to those in training. Note, though, that the performance on conditions matched to those in training is much more stable. Hence, when the model is developed to be used on matched data, a held-out portion of the training set can be used to select the best model.

A disadvantage of the proposed method over standard PLDA is that, currently, no proper way of handling trials with multiple enrollment samples is available. The released code can still handle such trials by computing the score as an average over the scores obtained comparing the test sample with each enrollment sample in the trial. Yet, this approach is likely suboptimal. In the future, we plan to work on an extension of the proposed method that would allow scoring of multi-sample enrollment trials using a probabilistic formulation, as for PLDA. Other plans for future work are to replace the PLDA stage by a heavy-tailed PLDA formulation which has been shown to be a better fit for the distribution of the x-vectors \citep{Silnova_HT}. Further, we will explore whether approaches like Bayesian model averaging can lead to more stable generalization performance across epochs. Finally, we will explore end-to-end training of the embedding extractor and the proposed backend.

\appendix
\section{Embedding Extractor Model}
\label{sec:appendix_embeddings}
The speaker embeddings used for the experiments in this work are standard x-vectors \citep{snyder2016deep,KaldiRecipe17}. These vectors are obtained as the pre-activations from a DNN trained to classify the speakers in the training data. 
The input features for the embedding extraction network are power-normalized cepstral coefficients (PNCC)~\citep{kim:12} which, in our experiments, gave better results than the more standard mel frequency cepstral coefficients (MFCCs). We extract 30 PNCCs with a bandwidth going from 100 to 7600 Hz and root compression of 1/15. The features are mean and variance normalized over a rolling window of 3 seconds. 

Non-speech frames are discarded using a DNN-based speech activity detection system (SAD). Frames with a log-posterior for the speech class above -0.5 are used for the extraction of embeddings. The SAD DNN has two hidden layers with 500 and 100 nodes, respectively, and uses 19-dimensional mel-frequency cepstral coefficients (MFCC) features (excluding $C_0$), stacked over a window of 31 frames around each frame, and mean and variance normalized over a 3 second rolling window. The model was trained on clean telephone and microphone data from a random selection of files from Mixer datasets (2004-2010), Fisher, and Switchboard. A 5 minute DTMF tone, and a selection of noise and music samples with and without speech added were included in the pool of training data.
The ground truth SAD labels needed for DNN training where generated by decoding training audio using an English senone DNN. The speech/non-speech labels were then produced by applying a threshold of 0.2 to the sum of the three silence senones posteriors from the DNN. For all augmented system training data, the SAD alignments from the raw audio were used as ground truth. 

The embedding extractor was trained with 234K signals from 14,630 speakers. This data was compiled from Switchboard, NIST SRE 2004 through 2008, NIST SRE 2012, Mixer6, Voxceleb1, and Voxceleb2 (train set) data. Voxceleb1 data had 60 speakers removed that overlapped with Speakers in the Wild (SITW). More details on these datasets are given in \ref{sec:appendix_backend_data}. All waveforms were up- or down-sampled to 16 KHz before further processing.  In addition, we down-sampled any data originally of 16 kHz or higher sampling rate (74K files) to 8 kHz before up-sampling back to 16 kHz, keeping two ``raw'' versions of each of these waveforms. This procedure allowed the embeddings system to operate well in both 8kHz and 16kHz bandwidths.

Augmentation of data was applied using four categories of degradations as in~\citep{mclaren:odyssey18}, including music and noise, both at 10 to 25 dB signal-to-noise ratio, compression, and low levels of reverb. We used 412 noises compiled from both freesound.org and the MUSAN corpus. Music degradations were sourced from 645 files from MUSAN and 99 instrumental pieces purchased from Amazon music. For reverberation, examples were collected from 47 real impulse responses available on echothief.com and 400 low-level reverb signals sourced from MUSAN. Compression was applied using 32 different codec-bitrate combinations with open source tools. We augmented the raw training data to produce 2 copies per file per degradation type (randomly selecting the specific degradation and SNR level, when appropriate) such that the data available for training was 9-fold the amount of raw samples. In total, this resulted in 2,778K files for training the speaker embedding DNNs. 

The architecture of our embeddings extractor DNN follows the Kaldi recipe~\citep{KaldiRecipe17}. The DNN is implemented in Tensorflow, trained using an Adam optimizer with chunks of speech between 2.0 and 3.5 seconds. Overall, we extract about 4K chunks of speech from each of the speakers. 
DNNs were trained over 4 epochs over the data using a mini batch size of 96 examples. 
We used dropout with a probability linearly increasing from 0.0 up to 0.1 at 1.5 epochs then linearly decreasing back to 0.0 at the final iteration. The learning rate started at 0.0005, increasing linearly  after 0.3 epochs reaching 0.03 at the final iteration while training simultaneously using 8 GPUs, averaging the parameters from the 8 jobs every 100 mini-batches.

\section{Details on Backend Training Data}
\label{sec:appendix_backend_data}
The training data for the different backends explored in this work is grouped in six different domains, given by:
\begin{itemize}[leftmargin=1.5em]\itemsep0em 
\item{\bf VOX}: The same Voxceleb 1 and 2 \citep{voxceleb} data used for embedding extractor training. It includes 7129 speakers. Recordings are interviews mostly, though not exclusively, in English. We exclude the test part of Voxceleb 2 for use in evaluation.
\item{\bf SWB}: Also included in embedding extractor training. It consists of 2389 speakers speaking through a telephone channel \citep{godfrey:92}.
\item{\bf MIX}: This includes all the English-only speakers from SRE datasets from 2004 through 2012 \citep{nist_sre0406,nist_sre08,nist_sre10,nist_sre12,greenberg2020} and Mixer6 \citep{mixer6} datasets used for embedding extractor training. It includes 2713 speakers recorded over telephone and microphone channels.
\item{\bf MIX ML}: This domain includes any bilingual speakers from the SRE datasets used for embedding extractor training. It consists of 1322 speakers. This is the only training domain on which we can create cross-language trials since all other domains are either exclusively in English or include speakers speaking a single language each.
\item{\bf FVCAUS}: This domain is composed of interviews and conversational excerpts from 328 Australian English speakers from the forensic voice comparison dataset~\citep{morrison2015forensic,Morrison_2012}. Audio was recorded using close talking microphones resulting in extremely clean waveforms compared to the rest of the domains. This set is not included in the training data for the embedding extractor. 
\item{\bf RATS SRC}:  This domain is composed of telephone calls in five non-English languages from 286 speakers. We only used the source data (not re-transmitted) of the DARPA RATS program~\citep{ref:rats_set} for the SID task. This set is not included in the training data for the embedding extractor.
\end{itemize}

For all domains, signals for which no information about the recording session could be obtained and all speakers for which a single session was available were discarded. The inclusion of the last two sets in backend training is important for achieving better generalization performance since those two sets consist of conditions that are underrepresented in the embedding extractor training data: very clean high-quality microphone data and non-English data.

The number of speakers listed above exclude 50 speakers from each domain that are held-out as development sets. The training samples from the first four domains are augmented with the same procedure as for embedding training, except that only one degraded version of each original segment is created instead of 8 as in embedding extractor training, and that noise and music is added at 5dB level instead of 10-25dB. This decision was made very early on during development of our PLDA backend because we saw small but consistent gains in results from using this lower SNR level.

\section{Details on Test Data}
\label{sec:appendix_test_data}

The following datasets were used to test the performance of the models in this work.

\begin{itemize}[leftmargin=1.5em]\itemsep0em
\item {\bf Sets held-out from training}: These sets were created from 50 speakers from each of the six training domains described in \ref{sec:appendix_backend_data} which are held-out during training. 

\item {\bf SITW}: The Speakers in the Wild (SITW) dataset contains speech samples in English from open-source media~\citep{sitwdb} including naturally occurring noises, reverberation, codec, and channel variability. This data is similar in nature to the Voxceleb data. 

\item {\bf SRE16}: The NIST Speaker Recognition Evaluation (SRE) 2016 dataset \citep{nist_sre16} includes variability due to domain/channel and language. 
We present results on the non-English conversational telephone speech which is recorded over a wide variety of handset types. We use both the labeled development set, with speech in Cebuano and Mandarin, and the evaluation set, with speech in Tagalog and Cantonese.

\item {\bf SRE18}: We show results on the Call My Net 2 (CMN2) subset of the SRE’18 dataset \citep{nist_sre18}. This subset has similar characteristics to the SRE16 dataset with the exception of focusing on a different language, Tunisian Arabic, and including speech recorded using VOIP instead of just PSTN calls.

\item {\bf SRE19}: We use the NIST SRE 2019 conversational telephone speech (CTS) dataset \citep{nist_sre19} which consists of data from the CMN2 collect that was unused for SRE18. The spoken language was Tunisian Arabic and the collection made over varying telephony channels and in varying environments.

\item {\bf Voices}: For  this  corpus \citep{richey2018,Nandwana2020},  audio was recorded in furnished rooms with background noise played in conjunction with foreground speech selected from the LibriSpeech corpus using 12 different microphones. 

\item {\bf FBI}: The FBI evaluation corpus was supplied by the Federal Bureau  of  Investigation  (FBI)  and  consists  of  14  distinct conditions   including   same/cross-channel   and   same/cross-language  trials with a wide range of difficulties.  This set of corpora were selected by the FBI for calibration research to represent a very wide range of different conditions, collection sources, environments, languages, and channels.  Details on this dataset and the conditions included in our experiments can be found in \citep{Ferrer:aslp18}. 

\item {\bf Vox2-Tst}: We use the test split of the Voxceleb 2 dataset \citep{voxceleb}. This set is composed of interviews downloaded from YouTube. Language labels are not available for this data but it appears to be mostly, though not completely, in English. 

\item {\bf LASRS}: The corpus is composed of 100 bilingual speakers from each of three languages, Arabic, Korean and Spanish \citep{Beck2004}. Each speaker is recorded in two separate sessions speaking English and their native language using several recording devices.  

\item  {\bf FVCCMN}: Composed of interviews and conversational excerpts from over 68 female Chinese speakers from the forensic voice comparison dataset \citep{zhang2011forensic,Morrison_2012}. Recordings were made with high-quality lapel microphones. This corpus is similar to the FVCAUS one used for backend training, except that the language is different.
\end{itemize}

Table \ref{tab:dsets} shows the statistics for development and evaluation sets.
For SITW, SRE16, SRE18, SRE19 and Voices we use the 1-side enrollment trials defined with the datasets. These sets also have a corresponding development set defined in their releases. In this paper we only use the one for SRE16, which we use for development. We do not show results for the other development splits since we found that conclusions were very similar for each development split compared to the corresponding evaluation split.
For LASRS, FVCCMN and Vox2-Tst we create exhaustive trials, excluding same-session trials. 

\begin{table}[!t]
\centering
\vspace{0.1cm}
\caption{The development and evaluation datasets with number of speakers (spk) and target/impostor (tgt/imp) trial counts. }
\vspace{-0.1cm}
\footnotesize
\begin{tabular}{l|l|rrr}
	& &   \multicolumn{3}{c}{Eval Split} \\
	Use & Dataset             &  \#spk & \#tgt & \#imp  \\ 
	\midrule
\multirow{2}{*}{Dev}
	& FVCCMN     &  68 & 16.4k & 1.1m \\
	& SRE16-Dev  &  20 & 3.3k & 13.2k \\
	\midrule
\multirow{8}{*}{Eval}
	& SRE16-Eval & 201 & 27.8k & 1.4m \\
	& SRE18-Eval &  289 & 48.5k & 1.6m \\
	& SRE19-Eval & 195 & 54k & 10.5m \\
	& Voices-Eval & 100 & 36.4k & 3.6m \\
	& Vox2-Tst  & 118 & 117.6k & 11.9m \\
	& SITW-Eval &  180 & 3.7k &  0.7m \\
	& FBI        & 2426 & 7.5k & 3.5m\\
	& LASRS      &  333 & 41.0k & 4.8m \\
	\bottomrule
\end{tabular}
\label{tab:dsets}
\end{table}

\bibliographystyle{elsarticle-harv} 

\bibliography{all-short-no-crossrefs.bib}

\begin{thebibliography}{68}
\expandafter\ifx\csname natexlab\endcsname\relax\def\natexlab#1{#1}\fi
\providecommand{\url}[1]{\texttt{#1}}
\providecommand{\href}[2]{#2}
\providecommand{\path}[1]{#1}
\providecommand{\DOIprefix}{doi:}
\providecommand{\ArXivprefix}{arXiv:}
\providecommand{\URLprefix}{URL: }
\providecommand{\Pubmedprefix}{pmid:}
\providecommand{\doi}[1]{\href{http://dx.doi.org/#1}{\path{#1}}}
\providecommand{\Pubmed}[1]{\href{pmid:#1}{\path{#1}}}
\providecommand{\bibinfo}[2]{#2}
\ifx\xfnm\relax \def\xfnm[#1]{\unskip,\space#1}\fi
\bibitem[{Beck et~al.(2004)Beck, Schwartz and Nakasone}]{Beck2004}
\bibinfo{author}{Beck, S.D.}, \bibinfo{author}{Schwartz, R.},
  \bibinfo{author}{Nakasone, H.}, \bibinfo{year}{2004}.
\newblock \bibinfo{title}{A bilingual multi-modal voice corpus for language and
  speaker recognition {(LASR)} services}, in: \bibinfo{booktitle}{Proc.
  Odyssey-04}, \bibinfo{address}{Toledo, Spain}.
\bibitem[{Bishop(2006)}]{bishop:ml}
\bibinfo{author}{Bishop, C.M.}, \bibinfo{year}{2006}.
\newblock \bibinfo{title}{Pattern Recognition and Machine Learning}.
\newblock \bibinfo{publisher}{Springer}.
\bibitem[{Brandschain et~al.(2013)Brandschain, Graff, Walker and
  Cieri}]{mixer6}
\bibinfo{author}{Brandschain, L.}, \bibinfo{author}{Graff, D.},
  \bibinfo{author}{Walker, K.}, \bibinfo{author}{Cieri, C.},
  \bibinfo{year}{2013}.
\newblock \bibinfo{title}{Mixer 6 speech}.
\newblock
  \bibinfo{howpublished}{\url{https://catalog.ldc.upenn.edu/LDC2013S03}}.
\bibitem[{Br\"ummer(2008)}]{FocalBilinear}
\bibinfo{author}{Br\"ummer, N.}, \bibinfo{year}{2008}.
\newblock \bibinfo{title}{Focal bilinear toolkit}.
\newblock
  \bibinfo{howpublished}{\url{http://niko.brummer.googlepages.com/focalbilinear}}.
\bibitem[{Br\"ummer(2010a)}]{em4splda}
\bibinfo{author}{Br\"ummer, N.}, \bibinfo{year}{2010}a.
\newblock \bibinfo{title}{{EM} for simplified {PLDA}}.
\newblock \bibinfo{howpublished}{\url{https://sites.
  google.com/site/nikobrummer/EMforSPLDA.pdf}}.
\bibitem[{Br\"ummer(2010b)}]{brummer_thesis}
\bibinfo{author}{Br\"ummer, N.}, \bibinfo{year}{2010}b.
\newblock \bibinfo{title}{Measuring, Refining and Calibrating Speaker and
  Language Information Extracted from Speech}.
\newblock Ph.D. thesis. Stellenbosch University.
\bibitem[{Br\"ummer and {De Villiers}(2010)}]{Brummer:odyssey10}
\bibinfo{author}{Br\"ummer, N.}, \bibinfo{author}{{De Villiers}, E.},
  \bibinfo{year}{2010}.
\newblock \bibinfo{title}{The speaker partitioning problem}, in:
  \bibinfo{booktitle}{Proc. Odyssey-10}, \bibinfo{address}{Brno, Czech
  Republic}.
\bibitem[{Br\"ummer and Doddington(2013)}]{brummer2013likelihood}
\bibinfo{author}{Br\"ummer, N.}, \bibinfo{author}{Doddington, G.},
  \bibinfo{year}{2013}.
\newblock \bibinfo{title}{Likelihood-ratio calibration using prior-weighted
  proper scoring rules}, in: \bibinfo{booktitle}{Proc. Interspeech},
  \bibinfo{address}{Lyon, France}.
\bibitem[{Br\"ummer and du~Preez(2006)}]{Brummer:csl06}
\bibinfo{author}{Br\"ummer, N.}, \bibinfo{author}{du~Preez, J.},
  \bibinfo{year}{2006}.
\newblock \bibinfo{title}{Application independent evaluation of speaker
  detection}.
\newblock \bibinfo{journal}{Computer Speech and Language} \bibinfo{volume}{20}.
\bibitem[{Br\"ummer and du~Preez(2013)}]{brummer:pav}
\bibinfo{author}{Br\"ummer, N.}, \bibinfo{author}{du~Preez, J.},
  \bibinfo{year}{2013}.
\newblock \bibinfo{title}{The {PAV} algorithm optimizes binary proper scoring
  rules}.
\newblock
  \bibinfo{howpublished}{\url{https://sites.google.com/site/nikobrummer/pav_optimizes_rbpsr.pdf}}.
\bibitem[{Br\"ummer et~al.(2014)Br\"ummer, Swart and van
  Leeuwen}]{brummer2014comparison}
\bibinfo{author}{Br\"ummer, N.}, \bibinfo{author}{Swart, A.},
  \bibinfo{author}{van Leeuwen, D.}, \bibinfo{year}{2014}.
\newblock \bibinfo{title}{A comparison of linear and non-linear calibrations
  for speaker recognition}, in: \bibinfo{booktitle}{Proc. Odyssey-14},
  \bibinfo{address}{Joensuu, Finland}.
\bibitem[{Burget et~al.(2011)Burget, Plchot, Cumani, Glembek, Matejka and
  Br\"ummer}]{burget:icassp11}
\bibinfo{author}{Burget, L.}, \bibinfo{author}{Plchot, O.},
  \bibinfo{author}{Cumani, S.}, \bibinfo{author}{Glembek, O.},
  \bibinfo{author}{Matejka, P.}, \bibinfo{author}{Br\"ummer, N.},
  \bibinfo{year}{2011}.
\newblock \bibinfo{title}{Discriminatively trained probabilistic linear
  discriminant analysis for speaker verification}, in:
  \bibinfo{booktitle}{Proc. ICASSP}, \bibinfo{address}{Prague}.
\bibitem[{Cumani et~al.(2011)Cumani, Batzu, Colibro, Vair, Laface and
  Vasilakakis}]{cumani2011comparison}
\bibinfo{author}{Cumani, S.}, \bibinfo{author}{Batzu, P.D.},
  \bibinfo{author}{Colibro, D.}, \bibinfo{author}{Vair, C.},
  \bibinfo{author}{Laface, P.}, \bibinfo{author}{Vasilakakis, V.},
  \bibinfo{year}{2011}.
\newblock \bibinfo{title}{Comparison of speaker recognition approaches for real
  applications}, in: \bibinfo{booktitle}{Proc. Interspeech},
  \bibinfo{address}{Florence, Italy}.
\bibitem[{Cumani et~al.(2013)Cumani, Br{\"u}mmer, Burget, Laface, Plchot and
  Vasilakakis}]{cumani2013pairwise}
\bibinfo{author}{Cumani, S.}, \bibinfo{author}{Br{\"u}mmer, N.},
  \bibinfo{author}{Burget, L.}, \bibinfo{author}{Laface, P.},
  \bibinfo{author}{Plchot, O.}, \bibinfo{author}{Vasilakakis, V.},
  \bibinfo{year}{2013}.
\newblock \bibinfo{title}{Pairwise discriminative speaker verification in the
  i-vector space}.
\newblock \bibinfo{journal}{IEEE Transactions on Audio, Speech, and Language
  Processing} \bibinfo{volume}{21}.
\bibitem[{Dehak et~al.(2011)Dehak, Kenny, Dehak, Dumouchel and
  Ouellet}]{Dehak11}
\bibinfo{author}{Dehak, N.}, \bibinfo{author}{Kenny, P.},
  \bibinfo{author}{Dehak, R.}, \bibinfo{author}{Dumouchel, P.},
  \bibinfo{author}{Ouellet, P.}, \bibinfo{year}{2011}.
\newblock \bibinfo{title}{Front-end factor analysis for speaker verification}.
\newblock \bibinfo{journal}{IEEE Transactions on Audio, Speech, and Language
  Processing} \bibinfo{volume}{19}.
\bibitem[{Ferrer et~al.(2008)Ferrer, Graciarena, Zymnis and
  Shriberg}]{FerrerEtAl:icassp2008}
\bibinfo{author}{Ferrer, L.}, \bibinfo{author}{Graciarena, M.},
  \bibinfo{author}{Zymnis, A.}, \bibinfo{author}{Shriberg, E.},
  \bibinfo{year}{2008}.
\newblock \bibinfo{title}{System combination using auxiliary information for
  speaker verification}, in: \bibinfo{booktitle}{Proc. ICASSP},
  \bibinfo{address}{Las Vegas}.
\bibitem[{Ferrer and McLaren(2020a)}]{ferrer2020discriminative}
\bibinfo{author}{Ferrer, L.}, \bibinfo{author}{McLaren, M.},
  \bibinfo{year}{2020}a.
\newblock \bibinfo{title}{A discriminative condition-aware backend for speaker
  verification}, in: \bibinfo{booktitle}{Proc. of ICASSP 2020},
  \bibinfo{address}{Barcelona, Spain}.
\bibitem[{Ferrer and McLaren(2020b)}]{ferrer2020speaker}
\bibinfo{author}{Ferrer, L.}, \bibinfo{author}{McLaren, M.},
  \bibinfo{year}{2020}b.
\newblock \bibinfo{title}{A speaker verification backend for improved
  calibration performance across varying conditions}, in:
  \bibinfo{booktitle}{Proc. Odyssey-20}, \bibinfo{address}{Tokyo, Japan}.
\bibitem[{Ferrer et~al.(2019)Ferrer, Nandwana, McLaren, Castan and
  Lawson}]{Ferrer:aslp18}
\bibinfo{author}{Ferrer, L.}, \bibinfo{author}{Nandwana, M.K.},
  \bibinfo{author}{McLaren, M.}, \bibinfo{author}{Castan, D.},
  \bibinfo{author}{Lawson, A.}, \bibinfo{year}{2019}.
\newblock \bibinfo{title}{Toward fail-safe speaker recognition: Trial-based
  calibration with a reject option}.
\newblock \bibinfo{journal}{IEEE/ACM Trans. Audio Speech and Language
  Processing} \bibinfo{volume}{27}.
\bibitem[{Ferrer et~al.(2005)Ferrer, S{\"o}nmez and
  Kajarekar}]{FerrerEtAl:Eurospeech2005}
\bibinfo{author}{Ferrer, L.}, \bibinfo{author}{S{\"o}nmez, K.},
  \bibinfo{author}{Kajarekar, S.}, \bibinfo{year}{2005}.
\newblock \bibinfo{title}{Class-dependent score combination for speaker
  recognition}, in: \bibinfo{booktitle}{Proc. Interspeech},
  \bibinfo{address}{Lisbon}.
\bibitem[{Garcia-Romero and Espy-Wilson(2011)}]{romero:lennorm}
\bibinfo{author}{Garcia-Romero, D.}, \bibinfo{author}{Espy-Wilson, C.},
  \bibinfo{year}{2011}.
\newblock \bibinfo{title}{Analysis of i-vector length normalization in speaker
  recognition systems}, in: \bibinfo{booktitle}{Proc. Interspeech},
  \bibinfo{address}{Florence, Italy}.
\bibitem[{Garcia-Romero et~al.(2020)Garcia-Romero, Sell and
  McCree}]{magnet:odyssey2020}
\bibinfo{author}{Garcia-Romero, D.}, \bibinfo{author}{Sell, G.},
  \bibinfo{author}{McCree, A.}, \bibinfo{year}{2020}.
\newblock \bibinfo{title}{{MagNetO}: X-vector magnitude estimation network plus
  offset for improved speaker recognition}, in: \bibinfo{booktitle}{Proc.
  Odyssey-20}, \bibinfo{address}{Tokyo, Japan}.
\bibitem[{Gneiting and Raftery(2007)}]{properscoring}
\bibinfo{author}{Gneiting, T.}, \bibinfo{author}{Raftery, A.E.},
  \bibinfo{year}{2007}.
\newblock \bibinfo{title}{Strictly proper scoring rules, prediction, and
  estimation} \bibinfo{volume}{102}.
\bibitem[{Godfrey et~al.(1992)Godfrey, Holliman and McDaniel}]{godfrey:92}
\bibinfo{author}{Godfrey, J.}, \bibinfo{author}{Holliman, E.},
  \bibinfo{author}{McDaniel, J.}, \bibinfo{year}{1992}.
\newblock \bibinfo{title}{Switchboard: Telephone speech corpus for research and
  development}, in: \bibinfo{booktitle}{Proc. ICASSP}, \bibinfo{address}{San
  Francisco}.
\bibitem[{Greenberg et~al.(2020)Greenberg, Mason, Sadjadi and
  Reynolds}]{greenberg2020}
\bibinfo{author}{Greenberg, C.S.}, \bibinfo{author}{Mason, L.P.},
  \bibinfo{author}{Sadjadi, S.O.}, \bibinfo{author}{Reynolds, D.A.},
  \bibinfo{year}{2020}.
\newblock \bibinfo{title}{Two decades of speaker recognition evaluation at the
  national institute of standards and technology}.
\newblock \bibinfo{journal}{Computer Speech and Language} \bibinfo{volume}{60}.
\bibitem[{Greenberg et~al.(2013)Greenberg, Stanford, Martin, Yadagiri,
  Doddington, Godfrey and Hernandez-Cordero}]{nist_sre12}
\bibinfo{author}{Greenberg, C.S.}, \bibinfo{author}{Stanford, V.M.},
  \bibinfo{author}{Martin, A.F.}, \bibinfo{author}{Yadagiri, M.},
  \bibinfo{author}{Doddington, G.R.}, \bibinfo{author}{Godfrey, J.J.},
  \bibinfo{author}{Hernandez-Cordero, J.}, \bibinfo{year}{2013}.
\newblock \bibinfo{title}{The 2012 {NIST} speaker recognition evaluation}, in:
  \bibinfo{booktitle}{Proc. Interspeech}, \bibinfo{address}{Lyon, France}.
\bibitem[{Guo et~al.(2017)Guo, Pleiss, Sun and Weinberger}]{guo:17}
\bibinfo{author}{Guo, C.}, \bibinfo{author}{Pleiss, G.}, \bibinfo{author}{Sun,
  Y.}, \bibinfo{author}{Weinberger, K.Q.}, \bibinfo{year}{2017}.
\newblock \bibinfo{title}{On calibration of modern neural networks}, in:
  \bibinfo{booktitle}{Proc. of the 34th International Conference on Machine
  Learning}, \bibinfo{address}{Sydney, Australia}.
\bibitem[{Ioffe(2006)}]{Ioffe:2006}
\bibinfo{author}{Ioffe, S.}, \bibinfo{year}{2006}.
\newblock \bibinfo{title}{Probabilistic linear discriminant analysis}, in:
  \bibinfo{booktitle}{Proc. of the 9th European Conference on Computer Vision},
  \bibinfo{address}{Graz, Austria}.
\bibitem[{Karam et~al.(2011)Karam, Campbell and Dehak}]{karam2011towards}
\bibinfo{author}{Karam, Z.N.}, \bibinfo{author}{Campbell, W.M.},
  \bibinfo{author}{Dehak, N.}, \bibinfo{year}{2011}.
\newblock \bibinfo{title}{Towards reduced false-alarms using cohorts}, in:
  \bibinfo{booktitle}{Proc. ICASSP}, \bibinfo{address}{Prague}.
\bibitem[{Kenny(2010)}]{kenny:10}
\bibinfo{author}{Kenny, P.}, \bibinfo{year}{2010}.
\newblock \bibinfo{title}{Bayesian speaker verification with heavy-tailed
  priors}, in: \bibinfo{booktitle}{Proc. Odyssey-10}, \bibinfo{address}{Brno,
  Czech Republic}.
\newblock \bibinfo{note}{Keynote presentation}.
\bibitem[{Kenny et~al.(2007)Kenny, Boulianne, Ouellet and
  Dumouchel}]{Kenny:JFA}
\bibinfo{author}{Kenny, P.}, \bibinfo{author}{Boulianne, G.},
  \bibinfo{author}{Ouellet, P.}, \bibinfo{author}{Dumouchel, P.},
  \bibinfo{year}{2007}.
\newblock \bibinfo{title}{Joint factor analysis versus eigenchannels in speaker
  recognition}.
\newblock \bibinfo{journal}{IEEE Transactions on Audio, Speech, and Language
  Processing} .
\bibitem[{Kenny et~al.(2010)Kenny, Reynolds and Castaldo}]{kenny:2010}
\bibinfo{author}{Kenny, P.}, \bibinfo{author}{Reynolds, D.},
  \bibinfo{author}{Castaldo, F.}, \bibinfo{year}{2010}.
\newblock \bibinfo{title}{Diarization of telephone conversations using factor
  analysis}.
\newblock \bibinfo{journal}{IEEE Journal of Selected Topics in Signal
  Processing} .
\bibitem[{Kim and Stern(2012)}]{kim:12}
\bibinfo{author}{Kim, C.}, \bibinfo{author}{Stern, R.}, \bibinfo{year}{2012}.
\newblock \bibinfo{title}{Power-normalized cepstral coefficients ({PNCC}) for
  robust speech recognition}, in: \bibinfo{booktitle}{Proc. ICASSP},
  \bibinfo{address}{Kyoto}.
\bibitem[{Kingma and Ba(2015)}]{kingma2014adam}
\bibinfo{author}{Kingma, D.P.}, \bibinfo{author}{Ba, J.}, \bibinfo{year}{2015}.
\newblock \bibinfo{title}{Adam: A method for stochastic optimization}, in:
  \bibinfo{booktitle}{Proc. of ICLR}, \bibinfo{address}{San Diego}.
\bibitem[{van Leeuwen and Br{\"u}mmer(2013)}]{van2013distribution}
\bibinfo{author}{van Leeuwen, D.A.}, \bibinfo{author}{Br{\"u}mmer, N.},
  \bibinfo{year}{2013}.
\newblock \bibinfo{title}{The distribution of calibrated likelihood-ratios in
  speaker recognition}, in: \bibinfo{booktitle}{Proc. Interspeech},
  \bibinfo{address}{Lyon, France}.
\bibitem[{Mandasari et~al.(2015)Mandasari, Saeidi and van
  Leeuwen}]{mandasari2015quality}
\bibinfo{author}{Mandasari, M.I.}, \bibinfo{author}{Saeidi, R.},
  \bibinfo{author}{van Leeuwen, D.A.}, \bibinfo{year}{2015}.
\newblock \bibinfo{title}{Quality measures based calibration with duration and
  noise dependency for speaker recognition}.
\newblock \bibinfo{journal}{Speech Communication} \bibinfo{volume}{72}.
\bibitem[{Mandasari et~al.(2013)Mandasari, Saeidi, McLaren and van
  Leeuwen}]{mandasari2013quality}
\bibinfo{author}{Mandasari, M.I.}, \bibinfo{author}{Saeidi, R.},
  \bibinfo{author}{McLaren, M.}, \bibinfo{author}{van Leeuwen, D.A.},
  \bibinfo{year}{2013}.
\newblock \bibinfo{title}{Quality measure functions for calibration of speaker
  recognition systems in various duration conditions}.
\newblock \bibinfo{journal}{IEEE Transactions on Audio, Speech, and Language
  Processing} \bibinfo{volume}{21}.
\bibitem[{Martin and Greenberg(2009)}]{nist_sre08}
\bibinfo{author}{Martin, A.F.}, \bibinfo{author}{Greenberg, C.S.},
  \bibinfo{year}{2009}.
\newblock \bibinfo{title}{{NIST} 2008 speaker recongition evaluation:
  Performance across telephone and room microphone channels}, in:
  \bibinfo{booktitle}{Proc. Interspeech}, \bibinfo{address}{Brighton}.
\bibitem[{Martin and Greenberg(2010)}]{nist_sre10}
\bibinfo{author}{Martin, A.F.}, \bibinfo{author}{Greenberg, C.S.},
  \bibinfo{year}{2010}.
\newblock \bibinfo{title}{The {NIST} 2010 speaker recognition evaluation}, in:
  \bibinfo{booktitle}{Proc. Interspeech}, \bibinfo{address}{Makuhari, Japan}.
\bibitem[{Matejka et~al.(2017)Matejka, Novotn{\`y}, Plchot, Burget,
  Diez~S{\'a}nchez and Cernock{\`y}}]{matejka2017analysis}
\bibinfo{author}{Matejka, P.}, \bibinfo{author}{Novotn{\`y}, O.},
  \bibinfo{author}{Plchot, O.}, \bibinfo{author}{Burget, L.},
  \bibinfo{author}{Diez~S{\'a}nchez, M.}, \bibinfo{author}{Cernock{\`y}, J.},
  \bibinfo{year}{2017}.
\newblock \bibinfo{title}{Analysis of score normalization in multilingual
  speaker recognition.}, in: \bibinfo{booktitle}{Proc. Interspeech},
  \bibinfo{address}{Stockholm}.
\bibitem[{McLaren et~al.(2018)McLaren, Castan, Nandwana, Ferrer and
  Yilmaz}]{mclaren:odyssey18}
\bibinfo{author}{McLaren, M.}, \bibinfo{author}{Castan, D.},
  \bibinfo{author}{Nandwana, M.}, \bibinfo{author}{Ferrer, L.},
  \bibinfo{author}{Yilmaz, E.}, \bibinfo{year}{2018}.
\newblock \bibinfo{title}{How to train your speaker embeddings extractor}, in:
  \bibinfo{booktitle}{Proc. of Speaker Odyssey}, \bibinfo{address}{{Les Sables
  d'Olonne}, France}.
\bibitem[{McLaren et~al.(2016)McLaren, Ferrer, Castan and Lawson}]{sitwdb}
\bibinfo{author}{McLaren, M.}, \bibinfo{author}{Ferrer, L.},
  \bibinfo{author}{Castan, D.}, \bibinfo{author}{Lawson, A.},
  \bibinfo{year}{2016}.
\newblock \bibinfo{title}{The speakers in the wild {(SITW)} speaker recognition
  database}, in: \bibinfo{booktitle}{Proc. Interspeech}, \bibinfo{address}{San
  Francisco}.
\bibitem[{McLaren et~al.(2014)McLaren, Lawson, Ferrer, Scheffer and
  Lei}]{mclaren2014trial}
\bibinfo{author}{McLaren, M.}, \bibinfo{author}{Lawson, A.},
  \bibinfo{author}{Ferrer, L.}, \bibinfo{author}{Scheffer, N.},
  \bibinfo{author}{Lei, Y.}, \bibinfo{year}{2014}.
\newblock \bibinfo{title}{Trial-based calibration for speaker recognition in
  unseen conditions}, in: \bibinfo{booktitle}{Proc. Odyssey-14},
  \bibinfo{address}{Joensuu, Finland}.
\bibitem[{Morrison et~al.(2015)Morrison, Zhang and {Enzinger et.
  al}}]{morrison2015forensic}
\bibinfo{author}{Morrison, G.}, \bibinfo{author}{Zhang, C.},
  \bibinfo{author}{{Enzinger et. al}, E.}, \bibinfo{year}{2015}.
\newblock \bibinfo{title}{Forensic database of voice recordings of 500+
  australian english speakers}.
\newblock
  \bibinfo{howpublished}{\url{http://databases.forensic-voice-comparison.
  net}}.
\bibitem[{Morrison et~al.(2012)Morrison, Rose and Zhang}]{Morrison_2012}
\bibinfo{author}{Morrison, G.S.}, \bibinfo{author}{Rose, P.},
  \bibinfo{author}{Zhang, C.}, \bibinfo{year}{2012}.
\newblock \bibinfo{title}{Protocol for the collection of databases of
  recordings for forensic-voice-comparison research and practice}.
\newblock \bibinfo{journal}{Australian Journal of Forensic Sciences}
  \bibinfo{volume}{44}, \bibinfo{pages}{155--167}.
\newblock \URLprefix \url{https://doi.org/10.1080%2F00450618.2011.630412},
  \DOIprefix\doi{10.1080/00450618.2011.630412}.
\bibitem[{Nagrani et~al.(2020)Nagrani, Chung, Xie and Zisserman}]{voxceleb}
\bibinfo{author}{Nagrani, A.}, \bibinfo{author}{Chung, J.S.},
  \bibinfo{author}{Xie, W.}, \bibinfo{author}{Zisserman, A.},
  \bibinfo{year}{2020}.
\newblock \bibinfo{title}{Voxceleb: Large-scale speaker verification in the
  wild}.
\newblock \bibinfo{journal}{Computer Speech and Language} \bibinfo{volume}{60}.
\bibitem[{Nandwana et~al.(2019)Nandwana, Ferrer, McLaren, Castan and
  Lawson}]{nandwana2019is}
\bibinfo{author}{Nandwana, M.K.}, \bibinfo{author}{Ferrer, L.},
  \bibinfo{author}{McLaren, M.}, \bibinfo{author}{Castan, D.},
  \bibinfo{author}{Lawson, A.}, \bibinfo{year}{2019}.
\newblock \bibinfo{title}{Analysis of critical metadata factors for the
  calibration of speaker recognition systems}, in: \bibinfo{booktitle}{Proc.
  Interspeech}, \bibinfo{address}{Graz, Austria}.
\bibitem[{Nandwana et~al.(2020)Nandwana, Lomnitz, Richey, McLaren, Castan,
  Ferrer and Lawson}]{Nandwana2020}
\bibinfo{author}{Nandwana, M.K.}, \bibinfo{author}{Lomnitz, M.},
  \bibinfo{author}{Richey, C.}, \bibinfo{author}{McLaren, M.},
  \bibinfo{author}{Castan, D.}, \bibinfo{author}{Ferrer, L.},
  \bibinfo{author}{Lawson, A.}, \bibinfo{year}{2020}.
\newblock \bibinfo{title}{{The VOiCES from a Distance Challenge 2019: Analysis
  of Speaker Verification Results and Remaining Challenges}}, in:
  \bibinfo{booktitle}{Proc. Odyssey-20}, \bibinfo{address}{Tokyo, Japan}.
\bibitem[{Nautsch et~al.(2016)Nautsch, Saeidi, Rathgeb and
  Busch}]{nautsch2016robustness}
\bibinfo{author}{Nautsch, A.}, \bibinfo{author}{Saeidi, R.},
  \bibinfo{author}{Rathgeb, C.}, \bibinfo{author}{Busch, C.},
  \bibinfo{year}{2016}.
\newblock \bibinfo{title}{Robustness of quality-based score calibration of
  speaker recognition systems with respect to low-snr and short-duration
  conditions.}, in: \bibinfo{booktitle}{Proc. Odyssey-16},
  \bibinfo{address}{Bilbao, Spain}.
\bibitem[{Pezeshki et~al.(2020)Pezeshki, Kaba, Bengio, Courville, Precup and
  Lajoie}]{pezeshki2020gradient}
\bibinfo{author}{Pezeshki, M.}, \bibinfo{author}{Kaba, S.O.},
  \bibinfo{author}{Bengio, Y.}, \bibinfo{author}{Courville, A.},
  \bibinfo{author}{Precup, D.}, \bibinfo{author}{Lajoie, G.},
  \bibinfo{year}{2020}.
\newblock \bibinfo{title}{Gradient starvation: A learning proclivity in neural
  networks}.
\newblock \bibinfo{journal}{arXiv preprint arXiv:2011.09468} .
\bibitem[{Prince(2007)}]{prince:plda}
\bibinfo{author}{Prince, S.}, \bibinfo{year}{2007}.
\newblock \bibinfo{title}{Probabilistic linear discriminant analysis for
  inferences about identity}, in: \bibinfo{booktitle}{Proceedings of the
  International Conference on Computer Vision}.
\bibitem[{{Przybocki} et~al.(2007){Przybocki}, {Martin} and
  {Le}}]{nist_sre0406}
\bibinfo{author}{{Przybocki}, M.A.}, \bibinfo{author}{{Martin}, A.F.},
  \bibinfo{author}{{Le}, A.N.}, \bibinfo{year}{2007}.
\newblock \bibinfo{title}{{NIST} speaker recognition evaluations utilizing the
  mixer corpora - 2004, 2005, 2006}.
\newblock \bibinfo{journal}{IEEE Transactions on Audio, Speech, and Language
  Processing} \bibinfo{volume}{15}.
\newblock \DOIprefix\doi{10.1109/TASL.2007.902489}.
\bibitem[{Ramoji et~al.(2020)Ramoji, Krishnan and Ganapathy}]{ramoji2020neural}
\bibinfo{author}{Ramoji, S.}, \bibinfo{author}{Krishnan, P.},
  \bibinfo{author}{Ganapathy, S.}, \bibinfo{year}{2020}.
\newblock \bibinfo{title}{Neural {PLDA} modeling for end-to-end speaker
  verification}, in: \bibinfo{booktitle}{Proc. Interspeech},
  \bibinfo{address}{Shanghai, China}.
\bibitem[{Richey et~al.(2018)Richey, Barrios, Armstrong, Bartels, Franco,
  Graciarena, Lawson, Nandwana, Stauffer, van Hout, Gamble, Hetherly,
  Stephenson and Ni}]{richey2018}
\bibinfo{author}{Richey, C.}, \bibinfo{author}{Barrios, M.A.},
  \bibinfo{author}{Armstrong, Z.}, \bibinfo{author}{Bartels, C.},
  \bibinfo{author}{Franco, H.}, \bibinfo{author}{Graciarena, M.},
  \bibinfo{author}{Lawson, A.}, \bibinfo{author}{Nandwana, M.K.},
  \bibinfo{author}{Stauffer, A.R.}, \bibinfo{author}{van Hout, J.},
  \bibinfo{author}{Gamble, P.}, \bibinfo{author}{Hetherly, J.},
  \bibinfo{author}{Stephenson, C.}, \bibinfo{author}{Ni, K.},
  \bibinfo{year}{2018}.
\newblock \bibinfo{title}{Voices obscured in complex environmental settings
  {(VOICES)} corpus}, in: \bibinfo{booktitle}{Proc. Interspeech},
  \bibinfo{address}{Hyderabad, India}.
\bibitem[{Rohdin et~al.(2018)Rohdin, Silnova, Diez, Plchot, Matejka and
  Burget}]{Rohdin2018}
\bibinfo{author}{Rohdin, J.}, \bibinfo{author}{Silnova, A.},
  \bibinfo{author}{Diez, M.}, \bibinfo{author}{Plchot, O.},
  \bibinfo{author}{Matejka, P.}, \bibinfo{author}{Burget, L.},
  \bibinfo{year}{2018}.
\newblock \bibinfo{title}{End-to-end {DNN} based speaker recognition inspired
  by i-vector and {PLDA}}, in: \bibinfo{booktitle}{Proc. ICASSP},
  \bibinfo{address}{Calgary, Canada}.
\bibitem[{Sadjadi et~al.(2019)Sadjadi, Greenberg, Reynolds and
  Mason}]{nist_sre18}
\bibinfo{author}{Sadjadi, S.O.}, \bibinfo{author}{Greenberg, C.S.},
  \bibinfo{author}{Reynolds, D.A.}, \bibinfo{author}{Mason, L.},
  \bibinfo{year}{2019}.
\newblock \bibinfo{title}{The 2018 {NIST} speaker recognition evaluation}, in:
  \bibinfo{booktitle}{Proc. Interspeech}, \bibinfo{address}{Graz, Austria}.
\bibitem[{Sadjadi et~al.(2020)Sadjadi, Greenberg, Singer, Reynolds, Mason and
  Hernandez-Cordero}]{nist_sre19}
\bibinfo{author}{Sadjadi, S.O.}, \bibinfo{author}{Greenberg, C.S.},
  \bibinfo{author}{Singer, E.}, \bibinfo{author}{Reynolds, D.},
  \bibinfo{author}{Mason, L.}, \bibinfo{author}{Hernandez-Cordero, J.},
  \bibinfo{year}{2020}.
\newblock \bibinfo{title}{The 2019 {NIST} speaker recognition evaluation cts
  challenge}, in: \bibinfo{booktitle}{Proc. Odyssey-20},
  \bibinfo{address}{Tokyo, Japan}.
\bibitem[{Sadjadi et~al.(2017)Sadjadi, Kheyrkhah, Tong, Greenberg and
  Reynolds}]{nist_sre16}
\bibinfo{author}{Sadjadi, S.O.}, \bibinfo{author}{Kheyrkhah, T.},
  \bibinfo{author}{Tong, A.}, \bibinfo{author}{Greenberg, C.S.},
  \bibinfo{author}{Reynolds, D.A.}, \bibinfo{year}{2017}.
\newblock \bibinfo{title}{The 2016 {NIST} speaker recognition evaluation}, in:
  \bibinfo{booktitle}{Proc. Interspeech}, \bibinfo{address}{Stockholm}.
\bibitem[{Silnova et~al.(2018)Silnova, Br\"ummer, Garcia-Romero, Snyder and
  Burget}]{Silnova_HT}
\bibinfo{author}{Silnova, A.}, \bibinfo{author}{Br\"ummer, N.},
  \bibinfo{author}{Garcia-Romero, D.}, \bibinfo{author}{Snyder, D.},
  \bibinfo{author}{Burget, L.}, \bibinfo{year}{2018}.
\newblock \bibinfo{title}{Fast variational {B}ayes for heavy-tailed {PLDA}
  applied to i-vectors and x-vectors}, in: \bibinfo{booktitle}{Proc.
  Interspeech}, \bibinfo{address}{Hyderabad, India}.
\bibitem[{Sizov et~al.(2014)Sizov, Lee and Kinnunen}]{sizov2014}
\bibinfo{author}{Sizov, A.}, \bibinfo{author}{Lee, K.A.},
  \bibinfo{author}{Kinnunen, T.}, \bibinfo{year}{2014}.
\newblock \bibinfo{title}{Unifying probabilistic linear discriminant analysis
  variants in biometric authentication}, in: \bibinfo{booktitle}{Joint IAPR
  International Workshops on Statistical Techniques in Pattern Recognition
  (SPR) and Structural and Syntactic Pattern Recognition (SSPR)},
  \bibinfo{organization}{Springer}. pp. \bibinfo{pages}{464--475}.
\bibitem[{Snyder(2017)}]{KaldiRecipe17}
\bibinfo{author}{Snyder, D.}, \bibinfo{year}{2017}.
\newblock \bibinfo{title}{{NIST SRE} 2016 xvector recipe}.
\newblock
  \bibinfo{howpublished}{\url{https://david-ryan-snyder.github.io/2017/10/04/model_sre16_v2.html}}.
\bibitem[{Snyder et~al.(2016)Snyder, Ghahremani, Povey, Garcia-Romero, Carmiel
  and Khudanpur}]{snyder2016deep}
\bibinfo{author}{Snyder, D.}, \bibinfo{author}{Ghahremani, P.},
  \bibinfo{author}{Povey, D.}, \bibinfo{author}{Garcia-Romero, D.},
  \bibinfo{author}{Carmiel, Y.}, \bibinfo{author}{Khudanpur, S.},
  \bibinfo{year}{2016}.
\newblock \bibinfo{title}{Deep neural network-based speaker embeddings for
  end-to-end speaker verification}, in: \bibinfo{booktitle}{Proc. of Spoken
  Language Technology Workshop (SLT)}.
\bibitem[{Solewicz and Koppel(2005)}]{Solewicz:Eurospeech2005}
\bibinfo{author}{Solewicz, Y.}, \bibinfo{author}{Koppel, M.},
  \bibinfo{year}{2005}.
\newblock \bibinfo{title}{Considering speech quality in speaker verification
  fusion}, in: \bibinfo{booktitle}{Proc. Interspeech},
  \bibinfo{address}{Lisbon}.
\bibitem[{Solewicz and Koppel(2007)}]{solewicz:ASLP07}
\bibinfo{author}{Solewicz, Y.}, \bibinfo{author}{Koppel, M.},
  \bibinfo{year}{2007}.
\newblock \bibinfo{title}{Using post-classifiers to enhance fusion of low- and
  high-level speaker recognition}.
\newblock \bibinfo{journal}{IEEE Transactions on Audio, Speech, and Language
  Processing} \bibinfo{volume}{15}.
\bibitem[{Tan and Mak(2017)}]{tan2017vector}
\bibinfo{author}{Tan, Z.}, \bibinfo{author}{Mak, M.W.}, \bibinfo{year}{2017}.
\newblock \bibinfo{title}{{i-Vector} {DNN} scoring and calibration for noise
  robust speaker verification.}, in: \bibinfo{booktitle}{INTERSPEECH},
  \bibinfo{address}{Stockholm}.
\bibitem[{Van~Leeuwen and Br{\"u}mmer(2007)}]{van2007introduction}
\bibinfo{author}{Van~Leeuwen, D.A.}, \bibinfo{author}{Br{\"u}mmer, N.},
  \bibinfo{year}{2007}.
\newblock \bibinfo{title}{An introduction to application-independent evaluation
  of speaker recognition systems}, in: \bibinfo{booktitle}{Speaker
  classification I: Fundamentals, Features, and Methods}.
  \bibinfo{publisher}{Springer-Verlag}.
\bibitem[{Walker and Strassel(2012)}]{ref:rats_set}
\bibinfo{author}{Walker, K.}, \bibinfo{author}{Strassel, S.},
  \bibinfo{year}{2012}.
\newblock \bibinfo{title}{The {RATS} radio traffic collection system}, in:
  \bibinfo{booktitle}{Proc. Odyssey-12}, \bibinfo{address}{Singapore}.
\bibitem[{Zhang and Morrison(2011)}]{zhang2011forensic}
\bibinfo{author}{Zhang, C.}, \bibinfo{author}{Morrison, G.},
  \bibinfo{year}{2011}.
\newblock \bibinfo{title}{Forensic database of audio recordings of 68 female
  speakers of standard chinese}.
\newblock
  \bibinfo{howpublished}{\url{http://databases.forensic-voice-comparison.net}}.

\end{thebibliography}

\end{document}